\documentstyle[prd,aps,preprint,tighten,epsf]{revtex}

\def\ch{{\cal H}}

\def\bx{{\bf x}}
\def\by{{\bf y}}

\def\bp{{\bf p}}

\newcommand{\Tr}{{\rm Tr}\,}
\renewcommand{\Re}{ {\rm Re}\, }

\def\non{\nonumber}
\def\beq{\begin{equation}}
\def\eeq{\end{equation}}
\def\beqa{\begin{eqnarray*}}
\def\eeqa{\end{eqnarray*}}
\def\beqs{\begin{eqnarray}}
\def\eeqs{\end{eqnarray}}

\begin{document}

\bibliographystyle{apsrev}

\preprint{CU-TP-978}

\title{{Heavy Quarks on Anisotropic Lattices:} \\ 
	{The Charmonium Spectrum}}

\author{Ping Chen \thanks{pchen@phys.columbia.edu}}

\address{Department of Physics, 
Columbia University, 
New York, NY10027, 
USA}

\date{May 18, 2000}

\maketitle

\begin{abstract}
We present results for the mass spectrum of $c{\bar c}$ mesons 
simulated on anisotropic lattices 
where the temporal spacing $a_t$ is only half of the spatial spacing $a_s$. 
The lattice QCD action is the Wilson gauge action 
plus the clover-improved Wilson fermion action.
The two clover coefficients 
on an anisotropic lattice are estimated using mean links 
in Landau gauge. 
The bare velocity of light $\nu_t$ has been tuned to keep
the anisotropic, heavy-quark Wilson action relativistic.
Local meson operators and three box sources are used in obtaining 
clear statistics for the lowest lying and first excited 
charmonium states of $^1S_0$, $^3S_1$, $^1P_1$, $^3P_0$ and $^3P_1$. 
The continuum limit is discussed by extrapolating from quenched simulations 
at four lattice spacings in the range 0.1 - 0.3 fm. 
Results are compared with the observed values in nature and
other lattice approaches. 
Finite volume effects and dispersion relations are checked.
\end{abstract}


\section{Introduction}

Lattice Quantum Chromodynamics (QCD) opens a gateway to the study 
of non-perturbative phenomena in the strong interaction world. 
To explain (and in some cases predict) why the elementary particles 
are as heavy as they are is not only exciting, 
it is also unavoidable in the validation of QCD as 
the standard model of the strong interactions. 
Unfortunately the lattice simulations are by no means cheap, 
it is typical for a project to take months or years to finish on 
the present fastest supercomputers.

The study of heavy quarks demands even more computing resources
than that of light quarks, 
while heavy quarks may be more interesting.
As standard lattice actions break down when the lattice spacing 
$a > \frac{1}{m_0}$, 
where $m_0$ is the bare quark mass, 
the imposition of a fine enough lattice spacing makes 
the studies of heavy quarks completely out of reach given
current computing power.
To treat heavy quarks, special lattice actions must be designed. 
The two dominant approaches are non-relativistic lattice QCD (NRQCD)  
and the heavy relativistic or Fermilab approach,
and there is the newer anisotropic relativistic approach used in our work.

The NRQCD approach \cite{Lepage,Davies} attempts to describe 
an effective field theory at low energy. 
Essentially the action is expanded in powers of the lattice spacing $a$,
as standard lattice actions are, 
and in powers of the heavy quark velocity $v^2$.
Practically the NRQCD method works well for the 
spin-independent $b{\bar b}$ system made of bottom quarks.
However, continuum extrapolation is impossible in NRQCD as 
the non-relativistic expansion requires $a m_0 > 1$. 
Also, to study spin splittings, 
higher order terms have to be added to the action. 
For the $c{\bar c}$ (charmonium) system,
present evidence \cite{NRQCD_SS1,NRQCD_SS2} 
suggests that the NRQCD approach breaks down.
The spin splittings in the charmonium spectrum do not converge 
when higher relativistic corrections are added to the non-relativistic action, 
or when quantum corrections are switched from one prescription to another.

The heavy-relativistic or Fermilab approach \cite{Fermilab} 
incorporates interactions from both the small- and large-mass limits. 
For heavy quarks, 
the lattice action can be interpreted in a non-relativistic light.
Yet as $m_0 a \rightarrow 0$, 
the action conforms exactly to the standard Wilson action for light quarks.
This is accomplished without any constraint on the value of $a m_0$, 
in contrast to $a m_0 > 1$ in NRQCD and 
$a m_0 \ll 1$ in Wilson action.
The Fermilab approach connects both ends smoothly.
Concretely, 
its lattice action up to $O(a^2)$ lattice errors is simply 
the standard clover-term improved Wilson action {\em without} 
imposing space-time exchange symmetry. 
The coefficients in front of the covariant derivatives and 
the clover term improvement now appear in two copies, a temporal one and 
a spatial one.
The difficulty is that,  
to achieve the elimination of $O(a)$  lattice artifact for heavy quarks, 
these coefficients must all be all mass dependent. 

The anisotropic relativistic approach \cite{KlassenQuark1,KlassenQuark2} 
goes one 
step further.
As in the heavy relativistic approach, 
the space-time exchange symmetry is not imposed on the lattice action.
The key difference is,
here in the anisotropic relativistic approach, 
the lattice itself is discretized differently along the temporal and 
spatial directions 
with the temporal spacing $a_t$ chosen finer 
than the spatial spacing $a_s$. 
Since no further symmetry besides the space-time exchange symmetry is broken, 
the anisotropic action has the same terms 
as the heavy relativistic action does.
Defining the true anisotropy $\xi = \frac{a_s}{a_t}$ as
the ratio of the spatial to the temporal spacing, 
we may consider the heavy relativistic approach as 
the $\xi = 1$ special case of the anisotropic relativistic approach.
In both approaches, the relativity of the lattice action is restored 
by a required tuning of the bare parameters in the action.

There is one obvious benefit of moving to an anisotropic lattice:
heavy meson propagators often die out very fast 
and thus leave too few time slices which are useful for mass fitting.
With a finer temporal lattice spacing,
this problem may be cured at relatively low cost.
Another equally important benefit is that, 
with $a_t m_0 \ll 1$ on an anisotropic lattice, 
the mass dependence of the improvement coefficients can be expected to 
be weaker, or absent all together as is the case 
for some coefficients classically.
Since the numerically determined clover coefficients 
deviate considerably from their perturbative estimates, 
their possible weak mass dependence may allow us to avoid a 
difficult, non-perturbative numerical determinations.

To be self-contained, now we review the theoretical framework 
laid out in \cite{KlassenQuark1,KlassenQuark2}.

\section{The anisotropic Wilson QCD action}

The anisotropic QCD action is the sum of the gauge action $S_G^{\xi}$
and the fermion action $S_F^{\xi}$
\beq
\label{eq:asym_action}
S^{\xi} = S_G^{\xi} +  S_F^{\xi} \, .
\eeq

\subsection{The anisotropic gauge action $S_G^{\xi}$}

On an anisotropic lattice the gauge action becomes
\begin{eqnarray}
\label{eq:asym_gauge_action}
S_G^{\xi}
 &=& \frac{\beta}{N_c} \left[ \frac{1}{\xi_o}
     \sum_{x, s>s\prime} \Re \Tr \left[ 1-P_{ss\prime}(x) \right]
     + \xi_o \sum_{x, s} \Re \Tr \left[ 1-P_{st}(x) \right]
     \right]  
\end{eqnarray}
where $\xi_o$ is the bare anisotropy, which equals to the 
true anisotropy $\xi = \frac{a_s}{a_t}$ only at the classical level.
Note that the anisotropic $\beta$ is the geometric mean of the $\beta$'s 
along the temporal and spatial directions, 
thus it corresponds to a coarser spatial spacing and a finer temporal 
spacing than given by its isotropic equivalent of same value. 
Here $a_s$ and $a_t$ refer to the actual ``physical'' lattice 
spacings as computed by examining the propagation of physical 
particles over distances of many lattice units. 
Standard renormalization argument guarantee us that 
the resulting long distance physics predicted by the action in 
eq.(\ref{eq:asym_gauge_action}) will appear consistent with 
relativity after this anisotropic interpretation of lattice scales is 
adopted.

It is suggested that the true anisotropy $\xi$ be fixed during
the continuum extrapolation, for reasons shown later. 
All the computations are done at $\xi = 2$ in this work.
The bare anisotropy $\xi_0$ is tuned at each $\beta$ to keep $\xi$ the same
\cite{KlassenGauge}.

\subsection{A few standard definitions}

The covariant first- and 
second-order lattice derivatives $\nabla_\mu$ and $\Delta_\mu$ are
defined through their operations on 
the quark field $q(x)$
\beqa
 \nabla_\mu q(x)  & = & 
 {1\over 2a_\mu}\, \biggl[ U_\mu(x) q(x+\mu) - U_{-\mu}(x) 
           q(x-\mu)\biggr] \\ 
 \Delta_\mu q(x)  & = & 
 {1\over a_\mu^2} \, \biggl[ U_\mu(x) q(x+\mu) + U_{-\mu}(x) q(x-\mu)
                                       -2 q(x) \biggr]   \, .
\eeqa
Here we employ the notation 
$U_{-\mu}(x) \equiv U_\mu(x-\mu)^\dagger$ 
for the parallel transporter from $x$ to $x-\mu$.
The lattice spacing 4-vector
$a_{\mu} = \{a_t, a_s \vec{1} \}$
is introduced to simplify the formulae.

The Euclidean gamma matrices and the Dirac matrices $\sigma_{\mu\nu}$ 
are defined by 
\beq  
\gamma_{\mu}  =  \gamma_{\mu}^{\dagger} \:, \hskip1cm
\{ \gamma_{\mu}, \gamma_{\nu} \}  =  2 \delta_{\mu\nu}   \:, \hskip1cm     
\sigma_{\mu\nu}  =  \frac{i}{2}[\gamma_{\mu}, \gamma_{\nu}]    .
\eeq
The field tensor $F_{\mu\nu}(x)$ is defined by
\beqs 
4 Q_{\mu\nu}(x)  & = & 
                   U(x,\mu) U(x+\hat\mu,\nu) U^\dagger(x+\hat\nu,\mu) 
                   U^\dagger(x,\nu)  \nonumber +          \non \\
               & & U(x,\nu) U^\dagger(x-\hat\mu+\hat\nu,\mu) 
	           U^\dagger(x-\hat\mu,\nu) U(x-\hat\mu,\mu) \nonumber +
                                                          \non \\
	       & & U^\dagger(x-\hat\mu,\mu) U^\dagger(x-\hat\mu-\hat\nu,\nu) 
                   U(x-\hat\mu-\hat\nu,\mu) U(x-\hat\nu,\nu) \nonumber +
                                                          \non \\
               & & U^\dagger(x-\hat\nu,\nu) U(x-\hat\nu,\mu) 
		   U(x+\hat\mu-\hat\nu,\nu) U^\dagger(x,\mu)         
                                                          \non \\
F_{\mu\nu}(x)   & = & 
                   \frac{-i}{2a^2}[Q_{\mu\nu} - Q_{\mu\nu}^\dagger] .
\eeqs

\subsection{The anisotropic fermion action $S_F^{\xi}$}

Back on an isotropic lattice, the following terms make up 
the clover improved quark action 
\beq
	m_0, \; 
	 \!\not\! \nabla_\mu, \; 
	a  \!\not\! \Delta_\mu, \;
 a\frac{C_{sw}}{2} \sum_{\mu<\nu}  \sigma_{\mu\nu} F_{\mu\nu}
\eeq
in which the lattice spacing $a$ is introduced to keep these terms of same 
dimension.

For either an anisotropic lattice, or heavy quarks on an isotropic lattice,
the space-time exchange symmetry should not be imposed at the level
of the lattice action. 
Instead, in order to achieve a relativistic dispersion relation 
between energy and momentum,
the coefficients in front of 
spatial terms in the fermion action 
have to be different from those in front of temporal terms. 
Thus, the anisotropic quark action is expected to be simply the 
standard isotropic action in duplicates, one spatial copy
and one temporal copy, which is indeed exactly the final form we choose:
\beqs  \label{eq:aniso_quark_action}
S_F^{\xi}
  &=& a_t a_s^3  
      \sum_{x} \bar{q}(x)  \left[ 
      m_0 + 
      \nu_t [
      \gamma_t \bigtriangledown_t - 
      \frac{a_t}{2} \bigtriangleup_t ]  +
      \nu_s \sum_s [
      \gamma_s \bigtriangledown_s - 
      \frac{a_s}{2} \bigtriangleup_s ]   
      \right.                  \nonumber \\
  & & \left.
      - \frac{a_s}{2} [ 
      C_{\rm sw}^t  \sum_{s} \sigma_{ts} F_{ts} + 
      C_{\rm sw}^s  \sum_{s<s\prime} \sigma_{ss\prime} F_{ss\prime} ]
      \right] q(x)                   \nonumber \\
  &=& a_t a_s^3 
      \sum_{x} \bar{q}(x)  \left[ 
      m_0 + 
      \nu_t  \!\not\! D^{\rm Wilson}_t +
      \sum_s \nu_s  \!\not\! D^{\rm Wilson}_s 
      \right.                  \nonumber \\
  & & \left.
      - \frac{a_s}{2} [ 
      C_{\rm sw}^t  \sum_{s} \sigma_{ts} F_{ts} + 
      C_{\rm sw}^s  \sum_{s<s\prime} \sigma_{ss\prime} F_{ss\prime} ]
      \right] q(x)               \: .    
\eeqs
In the second line above, 
the first- and second-order lattice derivatives are combined
into the $r=1$ Wilson operators 
$D_{\mu}^{\rm Wilson} =  \nabla_\mu - {a_\mu \over 2} \gamma_\mu \Delta_\mu$.

There are more bare parameters than in a standard quark action.
The two clover coefficients are labelled by $C_{\rm sw}^t$ and 
$C_{\rm sw}^s$. 
The bare velocity of light $\nu_s$ is to be tuned to 
restore relativity on an anisotropic lattice, 
which could be equally well achieved by adjusting $\nu_t$.  
Indeed we need to vary only one of them as the two cases are simply related by 
rescaling the quark fields. 
For later convenience we keep both $\nu_s$ and $\nu_t$ here. 
In practice we will always choose one of them to be 1 and tune 
the other quantity via the dispersion relation 
between meson energy and momentum.
We will refer to these cases as ``$\nu_s$-tuning'' and ``$\nu_t$-tuning''.

All the quantities in eq.(\ref{eq:aniso_quark_action}) are dimensionful 
with hidden $a_s$ or $a_t$. 
In order to program the quark action, 
we find it convenient to rewrite it in dimensionless quantities, 
i.e. quantities with a hat.
With
$\hat{m_0} \equiv a_t m_0$,
$\hat{q} \equiv a_s^{\frac{3}{2}} q$, 
$\hat{\nabla}_{\mu}  \equiv a_{\mu} \nabla_{\mu}$,
$\hat{\Delta}_{\mu}  \equiv a_{\mu}^2 \Delta_{\mu}$,
$\hat{D}^{\rm Wilson}_{\mu}  \equiv a_{\mu} D^{\rm Wilson}_{\mu}$ and
$\hat{F}_{\mu\nu} \equiv a_{\mu}a_{\nu} F_{\mu\nu}$, the quark action reads
\beqs \label{eq:aniso_lattice_quark_action}
S_F^{\xi}
  &=& \sum_{x} \hat{\bar{q}}(x) \left[
      \hat{m_0} + 
      \nu_t  \!\not\! \hat{D}^{\rm Wilson}_t +
      \frac{\nu_s}{\xi_0}  \sum_s \!\not\! \hat{D}^{\rm Wilson}_s 
      - \frac{1}{2} [ 
      C_{\rm sw}^t  \sum_{s} \sigma_{ts} \hat{F}_{ts} + 
      \right.                  
      \nonumber \\
  & & \left.
      \frac{C_{\rm sw}^s}{\xi_0}  
      \sum_{s<s\prime} \sigma_{ss\prime} \hat{F}_{ss\prime} ]
      \right]  \hat{q}(x) .
\eeqs
Note that we choose to use $\xi_0$ instead of $\xi = \frac{a_s}{a_t}$ 
in the action. 
While $\xi_0 = \xi$ holds only classically, 
the another choice would just redefine $\nu_s$ and $C_{\rm sw}^s$.

\section{Classical improvement of the anisotropic action}

The gauge action is already correct up to $O(a^2)$, so 
we only need to improve the quark action by choosing the right values
for the bare parameters. 
Below are all the possible terms up to $O(a)$ in the quark matrix
\beqs    \label{eq:allterms}
 &	m_0, \; 
        \!\not\! \nabla_t (1 + O(a m_0)) ,  \; 
        \sum_{s} \!\not\! \nabla_s (1 + O(a m_0)) , \;       \\\nonumber
 &      a \Delta_t, \;
        a \sum_{s} \Delta_s, \;
        a \sum_{s} [\!\not\! \nabla_s, \!\not\! \nabla_t], \;        
        a \sum_{s} \{\!\not\! \nabla_s, \!\not\! \nabla_t \}, \;        
        a \sum_{s<s\prime} \{\!\not\! \nabla_s, \!\not\! \nabla_{s\prime} \}
        \, .
\eeqs
Note the term        
$ a \sum_{s} [\!\not\!\nabla_s, \!\not\!\nabla_t] $
never arises on an isotropic light-quark action due to space-time 
exchange symmetry, and the last two anti-commutation terms are 
simply the clover terms in different faces.
Also note the lattice spacing $a$ has been reintroduced here
to keep track of dimensions, 
referring to either $a_t$ or $a_s$. 
All the coefficients in front of these terms are 
possibly mass-dependent.

\subsection{Field redefinition}

On the classical level the simplest way to derive the on-shell
$O(a)$ improved anisotropic quark action is to relate it by 
a field redefinition to an action that has manifestly 
no $O(a)$ discretization errors. 
Since a field redefinition is just a change of variable in 
the path integral, on-shell quantities are not affected.
The Jacobian of a field transformation matters only at 
the quantum level, where, in the case at hand, 
its leading effect is to renormalize the gauge coupling and
perhaps the bare anisotropy.

We start with the naive fermion action which has no $O(a)$ 
discretization errors and 
whose bare quark mass is the continuum one $m_c$
\beq
{\bar q_c}(x) [m_c + \!\not\!\nabla] q_c(x) \: .
\eeq
Then we apply the field redefinition 
${\bar q_c} = {\bar q} {\bar \Omega}$, 
$ q_c =  q \Omega$ 
with
\beqs
\Omega & = & 1 + \frac{\Omega_m}{2} a_t m_c + 
		 \frac{\Omega_t}{2} a_t \!\not\!\nabla_t +
                 \frac{\Omega_s}{2} a_s \vec{\!\not\!\nabla} \\\nonumber
{\bar \Omega} & = &  1 + \frac{{\bar \Omega}_m}{2} a_t m_c + 
                         \frac{{\bar \Omega}_t}{2} a_t \!\not\!\nabla_t +
                         \frac{{\bar \Omega}_s}{2} a_s \vec{\!\not\!\nabla}
              \: ,
\eeqs
where
$\Omega_{m,t,s}$ and ${\bar \Omega}_{m,t,s}$ are six pure numbers, 
possibly mass-dependent. 

With all terms in eq.(\ref{eq:allterms}) showed up, 
the quark matrix in terms of the new fields
${\bar q}(x)$  and $q(x)$ reads
\beqs  \label{eq:field_redef}
&&{\bar \Omega} \: [m_c + \!\not\!\nabla] \: \Omega
								\\ \non
&&= ( 1 + a_t m_c \frac{{\bar \Omega}_m + \Omega_m}{2} )
                        \, m_c \: + 
    ( 1 + (a_t m_c \frac{{\bar \Omega}_m + \Omega_m}{2} +
           a_t m_c \frac{{\bar \Omega}_t + \Omega_t}{2}) 
    ) \, \!\not\!\nabla_t  \; + \;
     \\\non
&&\quad
    (1 + (a_t m_c \frac{{\bar \Omega}_m + \Omega_m}{2} +
          a_s m_c \frac{{\bar \Omega}_s + \Omega_s}{2} )
    )   \,   \vec{\!\not\!\nabla}   \, + 
	\\\non
&&\quad  
   (a_t \frac{{\bar \Omega}_t + \Omega_t}{2} ) \Delta_t \: + \:
   (a_s \frac{{\bar \Omega}_s + \Omega_s}{2} ) 
        ( \sum_s \Delta_s + 
        \sum_{s<s\prime} \{\!\not\! \nabla_s \, , \, \!\not\! \nabla_{s\prime} \}
        )  \: +
     \\\non
&&\quad
   (a_s \frac{{\bar \Omega}_s - \Omega_s}{4} -
         a_t \frac{{\bar \Omega}_t - \Omega_t}{4} )  \,
         [\!\not\! \vec{\nabla} \, , \, \!\not\! \nabla_t ]  \: + \:
   (a_s \frac{{\bar \Omega}_s + \Omega_s}{4} +
        a_t \frac{{\bar \Omega}_t + \Omega_t}{4} )  \,
     \{\!\not\!\vec{\nabla} \, , \, \!\not\! \nabla_t \}   \: .
\eeqs
A few words on the notation here: 
to avoid excessively complicated, rigorous expressions, 
we have simply pulled 
${\bar \Omega}_s$, $\Omega_s$ and $a_s$ 
outside the spatial summation $\sum_s$ to make it explicit that 
they are common factors to all the three spatial directions.

\subsection{The classical $O(a)$ estimates of bare parameters}

To arrive at the anisotropic quark action in 
eq.(\ref{eq:aniso_quark_action}), we first get rid of 
the expensive and meritless term          
$[\!\not\! \vec{\nabla} \, , \, \!\not\! \nabla_t ]$
in eq.(\ref{eq:field_redef}) by requiring
$\Omega_m = {\bar \Omega}_m$, 
$\Omega_t = {\bar \Omega}_t$ and
$\Omega_s = {\bar \Omega}_s$, 
which sets no limitation on the remaining terms. 
Now if we recall that the clover terms  
\beq
\sigma_{\mu\nu} F_{\mu\nu} =  
 \{\!\not\!\nabla_\mu \, , \, \!\not\!\nabla_\nu \}  \, ,
\eeq
the quark matrix eq.(\ref{eq:field_redef}) becomes 
\beqs \label{eq:quark_mat111}
{\bar \Omega} \: [m_c + \!\not\!\nabla] \: \Omega & = &
    ( 1 + a_t m_c \Omega_m)      \, m_c \: + 
    ( 1 + (a_t m_c \Omega_m + a_t m_c \Omega_t)) \, \!\not\!\nabla_t  \; + \;
	\\\non
& &
    (1 + (a_t m_c \Omega_m + a_s m_c \Omega_s)) \, \sum_s\!\not\!\nabla_s \, + 
    (a_t \Omega_t) \Delta_t \: + \:
	\\\non
& &
   (a_s \Omega_s) ( \sum_s \Delta_s +  
                    \sum_{s<s\prime} \sigma_{ss\prime} F_{ss\prime} )  \: +
   (a_s \frac{\Omega_s}{2} + a_t \frac{\Omega_t}{2} )  \,
                    \sum_{s} \sigma_{st} F_{st}  \: .
\eeqs

We are left with three free parameters 
$\Omega_m$, $\Omega_s$ and $\Omega_t$. 
In the most attractive scheme, 
which actually leads to our definition of the quark action $S^{\xi}_F$ 
in eq.(\ref{eq:aniso_quark_action}), 
$\Omega_m$ is adjusted so that one (and only one) of 
    $\nu_s$ or $\nu_t$ equals 1,
$\Omega_s$ and $\Omega_t$ are adjusted so that 
the first- and second-order lattice derivatives are combined
into the Wilson operators with the full projection property
$D_{\mu}^{\rm Wilson} =  \nabla_\mu - {a_\mu \over 2} \gamma_\mu \Delta_\mu$.

In the case of $\nu_s$-tuning, the three parameters are set as 
$\Omega_m = \frac{1}{2}$, $\Omega_t = - \frac{1}{2}$ and 
$\Omega_s = - \frac{1}{2} 
              \frac{1+\frac{1}{2}a_t m_c}{1+\frac{1}{2}a_s m_c}$, 
giving the quark matrix as
\beqs \label{eq:quark_mat222}
{\bar \Omega} \: [m_c + \!\not\!\nabla] \: \Omega & = &
    (1 + \frac{1}{2} a_t m_c) \, m_c \: + \: \!\not\! D_t^{\rm Wilson} \; + \;
    \frac{1+\frac{1}{2} a_t m_c}{1+\frac{1}{2}a_s m_c} 
    \: \sum_s \!\not\! D_s^{\rm Wilson} \; - \;
	\\\non
& &
   \frac{a_s}{2} \sum_{s<s\prime} \sigma_{ss\prime} F_{ss\prime}  \: -
   \frac{1}{2}(\frac{a_s}{2} + \frac{a_t}{2}) \: 
   \sum_{s} \sigma_{st} F_{st}  \: .
\eeqs
From eq.(\ref{eq:quark_mat222}) we can read off 
the classical estimates of the bare parameters in 
the anisotropic quark action eq.(\ref{eq:aniso_quark_action}) 
if that action is to have no $O(a)$ errors 
\beqs \label{eq:nu_s_tuning}
m_0 = m_c ( 1 + \frac{1}{2} a_t m_c), \:
\nu_s = \frac{1+\frac{1}{2} a_t m_c}{1+\frac{1}{2}a_s m_c}, \:
\nu_t = 1, \:
C_{\rm sw}^s = 1, \:
C_{\rm sw}^t = \frac{1}{2}(1 + \frac{a_t}{a_s}) \: .
\eeqs

In the case of $\nu_t$-tuning, the three parameters are set as 
$\Omega_m = \frac{1}{2}\frac{a_s}{a_t}$, $\Omega_s = - \frac{1}{2}$ and 
$\Omega_t = - \frac{1}{2} 
              \frac{1+\frac{1}{2}a_s m_c}{1+\frac{1}{2} a_t m_c}$. 
The classical estimates of the bare parameters are
\beqs \label{eq:nu_t_tuning}
m_0 = m_c ( 1 + \frac{1}{2}a_s m_c), \:
\nu_s = 1, \:
\nu_t = \frac{1+\frac{1}{2}a_s m_c}{1+\frac{1}{2} a_t m_c}, \:
C_{\rm sw}^s = 1, \:
C_{\rm sw}^t = \frac{1}{2}(1 + \frac{a_t}{a_s}) \: .
\eeqs

The parameters specified by eq.(\ref{eq:nu_s_tuning}) and 
eq.(\ref{eq:nu_t_tuning}) correspond to our two 
possible conventions for $\nu_s$ and $\nu_t$. 
Since these two conventions are connected through a simple scale 
transformation of the fermion field 
$q(x) \rightarrow \sqrt{\nu_t / \nu_s} \, q(x)$, 
the parameters in eq.(\ref{eq:nu_t_tuning}) should equal 
those in eq.(\ref{eq:nu_s_tuning}) multiplied by 
$\frac{\nu_t}{\nu_s} = \frac{1+\frac{1}{2}a_s m_c}{1+\frac{1}{2} a_t m_c}$.
Of course, this simple scaling holds only to $O(a)$, a limitation that
will be improved below for the clover coefficients
$C_{\rm sw}^t$ and $C_{\rm sw}^s$.

\subsection{Better estimates of the clover coefficients}  
\label{section:better}

Since the $O(a)$ dependence of the clover coefficients 
leads to only $O(a^2)$ errors of the action
and we only aim to remove the $O(a)$ lattice artifact,
when writing down the classical estimates 
of $C_{\rm sw}^t$ and $C_{\rm sw}^s$ 
in eq.(\ref{eq:quark_mat222})-(\ref{eq:nu_t_tuning})
we have neglected the $O(a)$ parts of the transformation
coefficients $\Omega_s$ and $\Omega_t$.
However, the neglect of $O(a^2)$ terms in this manner is not necessary, 
indeed the clover coefficients are better
expressed in terms of the bare velocity of light $\nu_t$ or $\nu_s$.
In this way, 
we can partially determine 
the mass dependence of the clover coefficients
and also resolve the contradiction that the clover coefficients given in 
eq.(\ref{eq:nu_s_tuning}) and eq.(\ref{eq:nu_t_tuning}) 
are not related by the factor $\frac{\nu_t}{\nu_s}$.

This statement is based on the following observation on 
the general form of the quark action in eq.(\ref{eq:quark_mat111}).
We see that the spatial clover term 
$\sum_{s<s\prime} \sigma_{ss\prime} F_{ss\prime}$
always comes together with the spatial Wilson term
$\sum_s \Delta_s$ 
and there is also a similar relation between the temporal ones. 
Therefore, 
with the same values of $\Omega_m$, $\Omega_m$ and $\Omega_t$ 
given in above paragraphs, 
we can give a more precise version of 
eq.(\ref{eq:nu_s_tuning})-(\ref{eq:nu_t_tuning}).
In the $\nu_s$-tuning case, 
the more precise classical estimates of the bare parameters 
\beqs \label{eq:nu_s_tuning_better}
m_0 = m_c ( 1 + \frac{1}{2} a_t m_c), \:
\nu_s = \frac{1+\frac{1}{2} a_t m_c}{1+\frac{1}{2}a_s m_c}, \:
\nu_t = 1, \:
C_{\rm sw}^s = \nu_s, \:
C_{\rm sw}^t = \frac{1}{2}(\nu_s + \frac{a_t}{a_s}) \: .
\eeqs
In the $\nu_t$-tuning case, the estimates are
\beqs \label{eq:nu_t_tuning_better}
m_0 = m_c ( 1 + \frac{1}{2}a_s m_c), \:
\nu_s = 1, \:
\nu_t = \frac{1+\frac{1}{2}a_s m_c}{1+\frac{1}{2} a_t m_c}, \:
C_{\rm sw}^s = 1, \:
C_{\rm sw}^t = \frac{1}{2}(1 + \nu_t \frac{a_t}{a_s}) \: .
\eeqs

\section{Computational procedure}

We have run quenched simulations at four values of the lattice spacings 
(see the input parameters listed in table \ref{tab:basic}). 
The mass spectrum in the continuum limit is then obtained by 
extrapolating the measurements at these finite lattice spacings to 
zero lattice spacing (see results in table \ref{tab:charmonium}). 

\subsection{Adapt existing isotropic software to anisotropy}

Fortunately it is trivial to modify existing isotropic software to 
simulate anisotropic lattices, at least for quenched calculations.
Our practice is to re-scale the temporal links so that 
the anisotropic lattice action appears like 
the standard isotropic action. 
For the gauge sector, 
the temporal links are multiplied by $\xi_0$, 
transforming the gauge action from 
eq.(\ref{eq:asym_gauge_action}) into 
\beqs
S_G^\xi & \stackrel{U_t \rightarrow \xi_o U_t}{=} &
      \frac{1}{\xi_o} \frac{\beta}{N_c} 
      \sum_{x, \mu > \nu} \Re \Tr [ 1 - P_{\mu \nu}(x)]  
      \: + \: {\rm const}
      \: .
\eeqs
For the fermion sector, the temporal links are multiplied by 
$\frac{\nu_t}{\nu_s} \xi_0$, 
transforming the quark action from 
eq.(\ref{eq:aniso_lattice_quark_action}) into 
\begin{eqnarray}
S_F^{\xi} & \stackrel{U_t \rightarrow \frac{\nu_t}{\nu_s} \xi_0 U_t}{=} &
      \sum_{x} \hat{\bar{q}}(x) \left[
      \hat{m_0} + 
      \frac{\nu_s}{\xi_0} \!\not\! \hat{D}^{\rm Wilson} -
	\right.
	\\\non
&&    \left.
      \frac{1}{2} [ 
      \frac{\nu_s^2}{\nu_t^2 \xi_0^2} 
      C_{\rm sw}^t  \sum_{s} \sigma_{ts} \hat{F}_{ts} + 
      \frac{C_{\rm sw}^s}{\xi_0}  
      \sum_{s<s\prime} \sigma_{ss\prime} \hat{F}_{ss\prime} ]
      \right]  \hat{q}(x) .
\end{eqnarray}
In this way, we can use existing heat-bath code to update links and 
existing Dirac operator $\!\not\!\hat{D}^{\rm Wilson}$ to 
measure mass spectrum, 
as long as the code does not assume the $SU(3)$ properties
of the gauge links, for example, to 
reconstruct third row of these links.
The two different scalings are not a problem in quenched simulations, 
although it may require some thought in full simulations 
as both $S_G^{\xi}$ and $S_F^{\xi}$ are used in combination to 
update field configurations.

It is a popular practice in lattice simulations to write the 
quark action in terms of $\kappa$ and thus separate local terms $A$
from nonlocal terms $\!\not\!D^{\rm Wilson}_{\rm n.l.}$
\beq
\!\not\!D^{\rm Wilson}_{\rm n.l.} q(x)
= \frac{1}{2 a_\mu} \sum_\mu [
       (\gamma_\mu - r) U_\mu(x) q(x+\mu) - 
       (\gamma_\mu + r) U_{-\mu}(x) q(x-\mu)]  \, .
\eeq
On anisotropic lattices, {\em after} the rescaling of temporal links 
$U_t \rightarrow \xi_0 \frac{\nu_t}{\nu_s} U_t$, 
the quark action can be rewritten as 
\beq
S_F^{\xi}   
\stackrel{U_t \rightarrow \frac{\xi_o \nu_t}{\nu_s} U_t}{=}
      \frac{1}{2\kappa^\xi}
      \sum_{x} \hat{\bar{\psi}}(x) 
      [A^\xi + 
      \frac{\nu_s}{\xi_o} 2 \kappa^\xi \hat{\!\not\!D}^{\rm Wilson}_{\rm n.l.}]
      \hat{\psi}(x)
\eeq
where the anisotropic $\kappa$ and $A$ are defined as
\beqs
\kappa^\xi 
& = & \frac{1}{2 \left[ \hat{m}_0+r(\nu_t + \frac{(d-1)\nu_s}{\xi_o})\right]}  
	, \: \mbox{ {\rm in this work} } r = 1
      \\
A^\xi
  &\stackrel{U_t \rightarrow \frac{\xi_o \nu_t}{\nu_s} U_t}{=}& 
      1 - \kappa^\xi \left[ 
     \frac{\nu_s^2}{\xi_o^2 \nu_t^2} C_{\rm sw}^t 
      \sum_{s}  \sigma_{st} \hat{F}_{st} +
      \frac{1}{\xi_0} C_{\rm sw}^s\sum_{s < s^\prime} 
      \sigma_{s s^\prime} \hat{F}_{s s^\prime} 
      \right]                                    \: .                         
\eeqs

\subsection{Set the lattice scale}

The heat bath algorithm for updating the gauge configurations is 
the standard Creutz method \cite{Creutz} extended by 
Kennedy-Pendleton \cite{Kennedy} and Cabibbo-Marinari 
\cite{Cabibbo1}.
Moving from an isotropic lattice to an anisotropic lattice, 
we need two bare parameters, 
namely the bare anisotropy $\xi_0$ and $\beta$,
to specify the gauge couplings needed to generate the gauge links.
In order to extrapolate the measurements to the continuum limit 
and to run simulations at reasonable physical volume, 
we need to know the lattice spacings $a_s$ and $a_t$ in physical units, 
or alternatively the spatial spacing $a_s$ and 
the renormalized (true) anisotropy $\xi = a_s / a_t$.
This work has been done in \cite{KlassenGauge} and 
\cite{KlassenScale1,KlassenScale2,KlassenScale3}.
The part of these earlier results used in our simulation are listed 
in table \ref{tab:scale}.
We now briefly describe their methods.

In our simulation the renormalized (true) anisotropy $\xi$ 
has been fixed at $\xi = 2$. 
The choice of $\xi$ to be an integer makes it most convenient
to determine the relationship between $\xi$ and $\xi_0$ at 
a given $\beta$ \cite{KlassenGauge}. 
Basically two static quark potentials are compared with each other. 
Both quark-antiquark pairs are propagating in the same spatial direction.
One pair is separated in a spatial direction 
(different from the propagating direction, of course) while 
the other is separated by twice (or other value of $\xi$) the 
number of lattice sites in the temporal direction. 
The bare anisotropy $\xi_0$ is tuned until the two static quark potentials 
become identical. 

It is also a good idea to be cautious by keeping 
the value of the renormalized anisotropy $\xi$ fixed 
during the course of taking the continuum limit. 
In this way the scaling property of the mass spectrum is the only issue here 
and 
we avoid the complication all together 
that lattice physics might behave differently 
at different anisotropy. 
Beyond this reason one can always take the continuum limit along 
some other smooth curve of true anisotropy $\xi$
varying with lattice spacing $a$. 

The lattice spacing was determined very accurately in terms of 
the Sommer scale $r_0$ 
\cite{KlassenScale1,KlassenScale2,KlassenScale3}. 
The scale $r_0$ is defined 
via the force between a heavy quark and antiquark 
\beq
	r_0^2 \, F(r_0) = 1.65 \, .
\eeq
The constant $1.65$ is chosen so that 
$r_0 = 0.50 \, {\rm fm}$ 
is an intermediate distance quantity. 
By contrast another often used dimensionful quantity, 
the string tension $\sigma$, is only asymptotically defined 
at long distance and thus suffers from inexact assumption of 
leading intermediate distance corrections. 
Although both the Sommer scale $r_0$ and the string tension $\sigma$ 
are calculated from the static quark potential, 
the available data are much better at intermediate distance. 
Therefore it is superior to use 
$r_0 = 0.50 \, {\rm fm}$ 
to attach physical units to lattice observables.

We should note that although all successful potential models closely agree on 
the value of $r_0$ to be $0.50 \, {\rm fm}$, 
it does bring some systematic errors to our simulation. 
We decide {\em not} to quote any unjustified estimate of the errors 
on $r_0 = 0.50 \, {\rm fm}$ and leave it open to the reader
as to how to treat the errors from 
this choice of length scale and the effect of quenching.
We should also note that this error comes in only at the very 
end of our calculations when the mass spectrum extrapolated 
to zero lattice spacing is written in the physical unit of MeV. 
Note, both $\frac{a_s}{r_0}$ and $\xi_0 = \xi_0(\xi, \beta)$ 
were determined to $1\%$ accuracy.
Therefore when extrapolating in the lattice spacing 
we may neglect the errors from $\frac{a_s}{r_0}$ and $\xi_0$
and worry about the errors on 
our mass measurements only.

\subsection{Estimate the clover term coefficients}

On each gauge configuration generated with the heat-bath algorithm, 
we invert the fermionic matrix using the conjugate gradient (CG) 
method \cite{CG} with preconditioning.  
We then measure all mesonic states that can be obtained from bilinear sources 
without derivative operators (using the later would lead to more noisy 
correlators), as shown in table \ref{tab:meson_op}. 

In addition to the bare quark mass, these measurements require 
three more input parameters in the fermionic action.
Two of them are the temporal and spatial clover coefficients
$C_{\rm sw}^t$ and $C_{\rm sw}^s$. 
Since a non-perturbative determination of the clover coefficients
(say using the Schr\"{o}dinger functional) is a daunting project, 
$C_{\rm sw}^t$ and $C_{\rm sw}^s$ are estimated using
tree-level tadpole improvement. 
There is empirical evidence that tree-level tadpole improvement 
achieves more than two-loop or even three-loop perturbative improvement does. 
At tree-level tadpole improvement one starts with the classical action 
and then renormalizes each gauge link by its ``mean value''
\beq
	U_\mu \rightarrow \frac{U_\mu}{u_\mu}        \, .
\eeq
Clearly the mean links $u_\mu = \{u_t, u_s, u_s, u_s\}$ 
can not be defined  in a gauge invariant manner.
The prescription to isolate the true gauge-independent tadpole
contribution is to minimally renormalize the gauge links 
by choosing a maximum definition of the mean links $u_\mu$. 

This maximization of mean links leads us to determine them in Landau gauge
since on the lattice, the Landau gauge condition 
$\partial_\mu A_\mu = 0$ 
is achieved by maximizing the functional 
\beq \label{eq:landau_gauge}
	F[U] = \sum_{x,\mu} \frac{1}{a_\mu^2} \Re \Tr U_\mu (x)  \, .
\eeq
However, there is one subtlety regarding the ratio $a_t / a_s$ of 
spatial and temporal lattice spacing: 
which anisotropy should be used in this gauge fixing process, 
the bare or renormalized one? 
We choose the bare one $\xi_0$ based on 
the following empirical observation 
\cite{KlassenGauge,KlassenQuark1,KlassenQuark2}. 
The tadpole improvement hypothesis says that 
the ratio $u_t/u_s$ gives a tree-level estimate of the renormalization 
of the anisotropy $\xi = \frac{u_t}{u_s} \xi_0$ 
(this relation will be used in the next paragraph 
to simplify eq.(\ref{eq:tadpole})).
It is found in \cite{KlassenQuark1,KlassenQuark2} 
that with the choice of $\xi_0$ 
the measured $u_t/u_s$ in Landau gauge 
agrees quite well (within 2\% or less) with the values measured 
non-perturbatively in \cite{KlassenGauge}. 

The tree-level estimate of the clover coefficients is 
given in eq.(\ref{eq:nu_s_tuning})-(\ref{eq:nu_t_tuning}) as 
\beqs \label{eq:tree_level_clover}
C_{\rm sw}^s = 1 \: ,  \hskip1cm
C_{\rm sw}^t = \frac{1}{2}\left(1+\frac{1}{\xi}\right) \: .
\eeqs
Now we work on the tadpole correction starting from 
the quark action 
\beqs            \label{eq:tadpole}
S_F^{\xi}    
  &=& \sum_{x} \hat{\bar{q}}(x) \left[
      \hat{m_0} + 
      \nu_t  \!\not\! \hat{D}^{\rm Wilson}_t +
      \frac{\nu_s}{\xi}  \sum_s \!\not\! \hat{D}^{\rm Wilson}_s -
      \right.                  
      \nonumber \\
  & & \left.
      \frac{1}{2} [ 
      C_{\rm sw}^t  \sum_{s} \sigma_{ts} \hat{F}_{ts} + 
      \frac{C_{\rm sw}^s}{\xi}  
      \sum_{s<s\prime} \sigma_{ss\prime} \hat{F}_{ss\prime} ]
      \right]  \hat{q}(x) .
      \nonumber \\
  &\stackrel{\rm tadpole}{\longrightarrow} & 
      \sum_{x} \hat{\bar{q}}(x) \left[
      \hat{m_0} + 
      \frac{1}{u_t} \nu_t  \!\not\! \hat{D}^{\rm Wilson}_t +
      \frac{1}{u_s} \frac{\nu_s}{\xi}  \sum_s \!\not\! \hat{D}^{\rm Wilson}_s -
      \right.                  
      \nonumber \\
  & & \left.
      \frac{1}{2} [ 
      \frac{1}{u_t^2 u_s^2} C_{\rm sw}^t  \sum_{s} \sigma_{ts} \hat{F}_{ts} + 
      \frac{1}{u_s^4} \frac{C_{\rm sw}^s}{\xi}  
      \sum_{s<s\prime} \sigma_{ss\prime} \hat{F}_{ss\prime} ]
      \right]  \hat{q}(x) .
      \nonumber \\
  &\stackrel{\xi = \xi_0 u_t / u_s}{\longrightarrow} & 
      \frac{1}{u_t} \sum_{x} \hat{\bar{q}}(x) \left[
      u_t \hat{m_0} + 
      \nu_t  \!\not\! \hat{D}^{\rm Wilson}_t +
      \frac{\nu_s}{\xi_0}  \sum_s \!\not\! \hat{D}^{\rm Wilson}_s -
      \right.                  
      \nonumber \\
  & & \left.
      \frac{1}{2} [ 
      \frac{1}{u_t u_s^2} C_{\rm sw}^t  \sum_{s} \sigma_{ts} \hat{F}_{ts} + 
      \frac{1}{u_s^3} \frac{C_{\rm sw}^s}{\xi_0}  
      \sum_{s<s\prime} \sigma_{ss\prime} \hat{F}_{ss\prime} ]
      \right]  \hat{q}(x) .
\eeqs
Comparing above with the action eq.(\ref{eq:aniso_lattice_quark_action}) 
we simulate, the tree-level tadpole estimate of the clover coefficients is
\beqs \label{eq:clover_term_estimate}
C_{\rm sw}^s = \frac{1}{u_s^3} \: ,  \hskip1cm
C_{\rm sw}^t = \frac{1}{2}\left(1+\frac{1}{\xi}\right)
	           \frac{1}{u_t u_s^2}    \hskip1cm  .
\eeqs

The values given in eq.(\ref{eq:clover_term_estimate})
are what we use in this work. 
However, in retrospect, we think it may be better {\em not} to 
base the tadpole improvement on 
eq.(\ref{eq:nu_s_tuning})-(\ref{eq:nu_t_tuning}) or 
eq.(\ref{eq:tree_level_clover}).
Rather, there is a better classical estimate given in 
eq.(\ref{eq:nu_s_tuning_better})-(\ref{eq:nu_t_tuning_better}). 
In the $\nu_t$-tuning, the difference between 
eq.(\ref{eq:nu_s_tuning})-(\ref{eq:nu_t_tuning}) and
eq.(\ref{eq:nu_s_tuning_better})-(\ref{eq:nu_t_tuning_better}) 
is small and perhaps negligible, but this doesn't seem to be the case 
for the $\nu_s$-tuning. We will comment more on this in 
section \ref{section:tuning}.

\subsection{Tune the quark mass $m_0$ and the bare velocity of light $\nu_t$}

The remaining bare parameters in the quark action are 
the bare velocity of light $\nu_t$ and the bare quark mass $m_0$. 
Before we explain how these two inputs are tuned, 
we need to define the effective velocity of light $c(\bp)$ first.
In terms of the energy $E(\bp)$, the mass $m = E(0)$ and 
the momentum $\bp$ of a meson, 
the dispersion relation has the form of 
\beq
 E^2(\bp) = m^2 + \bp^2 + O(\sum_s p_s^4)  
		+ O(\sum_{s s^\prime} p_s^2 p_{s^\prime}^2).
\eeq
The effective velocity of light $c(\bp)$ is then given by 
\beq
	c(\bp) = \xi \sqrt{\frac{a_t^2 E^2(\bp) - a_t^2 m^2}{\bp^2 a_s^2}}
\eeq
where the factor $\xi = a_s / a_t$ comes from the fact that 
the lattice energy is expressed in temporal spacing $a_t$ 
while the momentum $\bp$ is expressed in spatial spacing $a_s$.
The fact that $c(\bp) \neq 1$ is due to finite lattice spacing error.

The bare quark mass $m_0$ and the bare velocity of light $\nu_t$ 
are tuned simultaneously so that the spin average 1S meson mass 
equals its observed value
\beq
 \frac{1}{4} m(1^3S_0) + \frac{3}{4} m(1^3S_1)  
    \stackrel{!}{=} 3.067 \, {\rm GeV}
\eeq
and $c(0)=1$ for the pseudo-scalar meson $\eta_c$. 

We obtain $c(0)$ by extrapolation from the $c(\bp)$ 
of the pseudo-scalar meson $\eta_c$ at the two 
lowest, on-axis momenta
$\frac{2\pi}{L}(1, 0, 0)$ and $\frac{2\pi}{L}(2, 0, 0)$, 
assuming $c(\bp) - c(0) \, \propto \, \bp^2$
(we may also choose some other direction in momentum space, say
$\frac{2\pi}{L}(1, 1, 0)$ and $\frac{2\pi}{L}(2, 2, 0)$, 
but the $O(p^4)$ errors in $c(\bp)$ will be larger).
To increase statistics 
we always average over momenta that just differ by permutations 
of their components.
Precisely we use 
\beq
	c(0) = \sqrt{\frac{16 E(1)^2 - E(2)^2 - 15 m^2}
		{12 (2\pi/L)^2}}.
\eeq
where the three energies
$E(2) = E(\bp = \frac{2\pi}{L}(2,0,0))$, 
$E(1) = E(\bp = \frac{2\pi}{L}(1,0,0))$, 
and $m = E(\bp = \frac{2\pi}{L}(0,0,0))$
 are from a correlated fit of hadronic correlators at 
three momenta.
But this formula is merely one way of computing $c(\bp)$.
The point is 
we eliminate the leading relativistic errors completely 
by demanding $c(0)=1$ up to $O(p^4)$ error.

The tuning here involves two to four iterations to get both 
the quark mass $m_0$ and the bare velocity of light $\nu_t$ 
right to about $1\%$ (see table \ref{tab:basic}), 
an accuracy in line with that of scale setting $a_s/r_{0N}$ and $\xi_0$. 
Generally, the simultaneous tuning of more than one parameter to 
such a precision can be quite expensive, but here is relatively 
easy, since we find that the mass dependence of $\nu_t$ is 
very weak on anisotropic lattices, just as in the classical case. 
Besides it is sufficient to use about 100 configurations for the 
initial tuning. 
For the final measurement runs using the tuned parameters 
we have over 400 configurations per lattice spacing. 

In our experience,
an increase of the bare velocity of light 
$\nu_t$ leads to smaller meson masses
and effective velocity of light $c(0)$; 
not surprisingly, an increase of the bare quark mass $m_0$ leads to 
larger meson masses although 
such an increase has only very minor effect on $c(0)$.
Thus, one may tune the bare velocity of light $\nu_t$ first 
until $c(0) = 1$ and then work on the bare quark mass $m_0$ to 
make the meson mass $m(1S)$ right. 

Alternatively we may set $\nu_t = 1$ and tune $\nu_s$ instead. 
The best value of $\nu_s$ should be around the inverse of 
the best value of $\nu_t$. 
The desired 
$C_{\rm sw}^t$, $C_{\rm sw}^s$ and $m_0$ of $\nu_s$-tuning 
are simply 
their corresponding value of $\nu_t$-tuning divided by $\nu_t$
since two cases are related by rescaling the fields by 
a factor of $\nu_t$.

All together in this work we have studied the dispersion relation
using the six lowest momenta.
Omitting the common factor $\frac{2\pi}{L}$, they are
\beqs
\bp_{0..6} & = & [0, 0, 0],  [0, 0, 1],	[0, 0, 2],  
		     	[0, 1, 1],   	[0, 2, 2],  
		     	[1, 1, 1],   	[2, 2, 2].
\eeqs
While $\bp_3$ and $\bp_4$ are useful in double checking the tuning from 
$\bp_{0..2}$,  $\bp_5$ and $\bp_6$ are too noisy to be of any use.

\subsection{Fit hadronic correlators of local sink and three box sources}

Our computation gives a very extensive mass spectrum,
namely, the radial $n=1$ ground states and $n=2$ first excitations
of the $^1S_0$, $^3S_1$, $^1P_1$, $^3P_0$ and $^3P_1$ particles as
listed in table \ref{tab:meson_op}.
An anisotropic lattice certainly gives the benefit of fine 
temporal spacing without much computing expense.
Had the simulation been performed on an isotropic lattice of equivalent cost, 
the signal of the heavy hadronic correlators may 
have died out quickly, 
making the mass fitting technically impossible. 

On each gauge configuration we compute the quark propagator 
three times. Each time only the size of the box source differs, 
as detailed in table \ref{tab:sim_cnd}. 
Each calculation uses local sinks. 
Although this combination is largely due to software availability, 
it is sufficient to generate masses of the ground state and first excited state
with better precision than given by typical lattice calculations 
(see results in fig \ref{fig:charmonium} and table \ref{tab:charmonium}).
The trick is to gear the size of the box source to 
correspond to the desired wave-function size.
Fortunately we need to do this tuning at only one value of lattice spacing.
Since we know the lattice spacings in terms of physical units, 
we can then easily estimate the optimal box sizes for 
other values of lattice spacings.

\def\hadrcr{\bar{\ch}(\by, 0)}
\def\hadrde{{\ch}(\bx, \tau)}
\def\hadrop{\bar{\ch}}

In theory a fitting ansatz should incorporate 
the energies $E_1(\bp)$, $E_2(\bp)$, etc of each state
entering the hadronic correlator with
point source $\by$ and point sink $\bx$
\beqs  \label{eq:fitting_function}
    <\sum_{\bx} e^{i\frac{2\pi}{L} \bp \cdot \bx} \hadrde \hadrcr> 
      & = & 
	\sum_n\,|<n|\hadrop|0>|^2\,e^{-E_n(\bp)\tau} \, .
\eeqs
However, we only have a limited number of time slices
and therefore we want to reduce 
the number of fitting parameters as much as possible, so long as 
the fitting ansatz still closely reflects 
the time dependence of the underlying hadronic correlator. 

The three sizes of box sources at each lattice spacing 
will be referred as small-size, medium-size and large-size.
A 1-cosh fitting ansatz (i.e. ground state only) applies to 
the hadronic correlators with the medium-size box source so well that 
the fitted ground state mass stabilizes for a fitting range as early as 
the minimum time slice ${t_{\rm min}} = \frac{T}{8}$. 
The other two hadronic correlators are to be fitted with a 2-cosh ansatz
or a 3-cosh ansatz to give masses of excited states. 
We did not use hadron correlators with a point source because 
the undesirable contributions they receive from higher excited states 
(of radial quantum number $n \geq 3$) can not be overlooked. 
Instead, we speculate that a small-size 
box source behaves like a ``mild'' point source. 
Just as $|<n|\hadrop|0>|^2$ in the case of a point source, 
the contributions of each energy state in the hadronic correlator of 
a small box source are all positive. 
Nevertheless, the correlators containing a small-size box source do not 
appear to be contaminated by higher excited states beyond our interest.
Meanwhile the correlators of a large-size box source may behave 
like a wall source correlator. 
The amplitudes of different energy states 
may be a combination of positive and negative numbers.
When fitting three correlators simultaneously
it is indeed a desirable feature 
that a physical state manifests itself 
in these correlators with different signs.

Powell's method is used to minimize the $\chi^2$ to 
generate the best-fit parameters.
Developed by Kent Hornbostel and Peter Lepage, 
the fitting code supports an arbitrary (correlated) fitting ansatz 
on arbitrary numbers of data files. 
We use 1-cosh, 2-cosh and 3-cosh fitting ansatze. 
The $t_{\rm max}$ of a fitting range is typically fixed through 
the effective mass calculation
while the $t_{\rm min}$ is varied through all values as long as there are 
enough degrees of freedom left. 

To be selected, a fitted result ought to satisfy three criteria:
1. It is consistent with all other fitting ansatze.
   For example, although the 2-cosh ansatz reaches plateau earlier 
   than the 1-cosh ansatz does 
   and only one of them may succeed in generating 
   meaningful best-fit parameters, 
   the ground state mass should agree between both ansatz.
2. It becomes stable when $t_{\rm min}$ reaches certain value.
3. All the fitted quantities including amplitudes are statistically 
   nonzero and have the right sign if known. 
Precisely the fitting procedure is:
\begin{enumerate}
\item Calculate the effective mass on each adjacent pair of time 
      slices using the 1-cosh ansatz. 
      This step supplies the value of $t_{\rm max}$ 
      and the estimates of ground state masses and amplitudes
      to step 2. 
      It also gives us an opportunity to easily examine 
      the autocorrelation between configurations.
\item Apply the 1-cosh fitting ansatz on each individual datafile to 
      give better initial guesses for the real fittings to follow.
      One datafile corresponds to one unique combination of mesonic 
      operator $\bar{\psi}\Gamma\psi$, momentum \bp, and box source size.
\item The spin average mass $m(1S)$ and the mass differences 
      $\triangle m_{1^3S_1 - 1^1S_0}$, $\triangle m_{1P - 1S}$, 
      and $\triangle m_{1^3P_1 - 1^3P_0}$
      are obtained using the correlated 1-cosh ansatz on 
      two or three (i.e. the number of particles involved) 
      correlators at zero momentum, $\bp = 0$,  
      and medium-size box sources.
      Only the correlators of medium-size box sources are used here 
      since they are designed to suit the 1-cosh ansatz very well. 
      If we otherwise include correlators of small-size box sources and 
      large-size box sources, 
      we will have to add more states into the fitting ansatz.
      The fitting will then become too complex to succeed. 
      Since one of the triplet $P$-wave states, $1^3P_2$, is missing, 
      we do not have the true spin-average of the $1^3P_J$ states. 
      Instead we have used $1^1P_1$ in the $1^1P_1-1S$ splitting. 
\item To monitor the effective velocity of light $c(\bp = 0)$, 
      a correlated 1-cosh ansatz is applied on three $^1S_0$ correlators 
      of momentum 0, $\frac{2\pi}{L}(1, 0, 0)$ and $\frac{2\pi}{L}(2, 0, 0)$. 
      Again, only correlators of medium-size box sources are used here. 
      We observed however, that the higher the underlying momentum is, 
      the smaller would be the best medium-size source for a 1-cosh ansatz.
      Whatever size we choose for the box source, it will not make the
      correlators of all momenta simultaneously perfect for the 1-cosh fitting.
      But since the effect is weak and the fitted results are stable
      across different $t_{\rm min}$, we should not worry too much 
      about the relatively worse $\chi^2/{\rm d.o.f.}$ here.
\item To obtain the masses of the ground state and first excitation of 
      each particle $^{2S+1}L_J$, 
      the correlated 2-cosh ansatz is applied on its correlators 
      from small- and large-size box source 
      while a 1-cosh ansatz is used simultaneously for the medium-size 
      box source. 
      We always apply 1-cosh ansatz on the correlators with 
      medium-size box source because
      the excited state amplitude of correlators from 
      medium-size box sources are statistically zero when fitted with 
      2-cosh ansatz.
\item The 3-cosh fitting is done for the purpose of a sanity check.
\end{enumerate}

The masses and mass differences are quoted in
table \ref{tab:charmonium} and \ref{tab:splitting}. 
The fitting details such as $\chi^2/{\rm d.o.f.}$ are listed in 
table \ref{tab:fit_b5.6}-\ref{tab:fit_b6.1}.
In these tables the $Q$ value is a normalized indicator 
of the quality of a fit, 
defined as 
the probability that we would end up with a higher $\chi^2$
($\chi^2_{\rm min}$ if more strictly speaking) if we did the simulations
many times, it takes values between 0 and 1. 
The dropped eigenvalues refer to the truncated 
smallest eigenvalues of the sample correlation matrix.
The goodness-of-fit is chosen as the product of the $Q$ value and the 
degrees of freedom (d.o.f.). 
The d.o.f. in turn is the number of time slices from all 
correlators minus the number of fitting parameters and 
the number of dropped eigenvalues of the correlation matrix.

The error of a fitted quantity is given as 
the amount of perturbation away 
from the best-fitted value in order to increase the $\chi^2$ by 1. 
Assuming the data model is right, this definition indeed gives the 
68\% range of a fitted parameter \cite{Recipe}. 
And it is much faster than the bootstrap or jackknife methods 
because it avoids the labor of producing and fitting synthetic data sets.
For a poor fit, usually signaled by a large 
$\chi^2/{\rm d.o.f.} \geq 1.5$, 
this definition may underestimate the statistical error.
The majority of our fits have $\chi^2/{\rm d.o.f.} \sim 1$, 
thereby we expect the errors from $\triangle \chi^2 = 1$ are good estimates. 
It is also a good idea to look at 
the fluctuations of the fitted results under a variety of 
fitting conditions 
and then to take these fluctuations into account 
in quoting the statistical errors.

\subsection{Extrapolate to the continuum limit}

Ultimately we want to reach the continuum limit by extrapolating 
from the computations at four values of lattice spacings. 
Only then can we compare our results with, and in some cases predict,
the mass spectrum in Nature. 

The mass differences, not the masses themselves, are to be extrapolated 
to the continuum limit for several reasons: 
1. We have tuned the spin average meson mass m(1S) to equal
   their experimental value, 
   therefore the masses will more or less right by design.
2. The tuning is not perfect, but instead is accurate to 1\%. 
   Consequently the mass spectrum from the computation 
   at one lattice spacing 
   may all be systematically lifted up, 
   while from a different run 
   they may all be dragged down. 
   Such overall mass shifts cancel out in the mass differences.
3. The masses are not independent quantities since they are 
   all measured on the same gauge background. 
   A correlated fit for the mass difference may be more precise than 
   a naive subtraction of two masses fitted independently.

The computations at different lattice spacings are independent computations, 
which makes the continuum extrapolation technically very easy.
Both the Wilson gauge action and the clover-term improved 
Wilson fermion action are accurate to $O(a^2)$.
Therefore the measurements at four values of $a_s^2$  
are fitted with a straight line 
(see table \ref{tab:splitting} and 
figure \ref{fig:splitting0}-\ref{fig:splitting1}). 
At $\beta = 5.7$, the run on a $16^3 \cdot 64$ lattice is used 
due to its better fitting quality than that of the run on  
a $8^3 \cdot 32$ lattice.
The intercept at $a_s^2 = 0$ gives the continuum extrapolated value.
The extrapolation is done 
using the regression functionality of the software {\em Xmgr}, 
and is double checked using {\em Maple}.

\section{Results for the charmonium spectrum}

Now we present our results and compare them with (if available) 
the experimental data and other lattice approaches.

\subsection{Finite volume effects}
In this work we have run simulations on four values of lattice spacings.
As is important in every lattice calculation, 
we need to make sure the simulation volumes are large enough to 
avoid significant finite-volume effects 
yet also small enough to avoid high computational cost.
Among the four $\beta$ values, 
the $\beta = 6.1$ run has the finest lattice spacing, 
its lattice of $16^3 \cdot 64$ sites extends 1.536 fm along
the spatial directions.
Therefore we choose to check for the finite volume effects at 
$\beta = 5.7$ on two lattices $8^3 \cdot 32$ and $16^3 \cdot 64$, 
corresponding to a spatial extension of 1.658 fm and 3.315 fm 
respectively (see table \ref{tab:scale}).
As listed in tables \ref{tab:basic} and \ref{tab:sim_cnd}, 
the identical input parameters
are chosen for the two $\beta=5.7$ runs. 
The mean links in Landau Gauge are measured on a lattice of $16^3 \cdot 48$
for the purpose of estimating the clover term coefficients. 

Comparing the mass spectrum between these two volumes at $\beta = 5.7$
(see tables \ref{tab:charmonium} and \ref{tab:splitting}), 
we see no sign of an overall volume bias and 
for the radial $n = 1$ ground states we see 
broad agreement within one sigma. 
The fitting of the $16^3 \cdot 64$ run 
is easier and more stable (over $t_{\rm tmin}$) than 
the fitting of the $8^3 \cdot 32$ run.
The reason is unclear --- 
presumably at the larger volume each hadron correlator 
fluctuates less after the time slice average.
However, 
we do expect both fittings to work better if 
the smallest box source size 
$2 \cdot 2 \cdot 2$ was slightly larger, say $2 \cdot 3 \cdot 3$.

The two $c(0)$'s are consistent within 1\%
(see table \ref{tab:basic}). 
Note the bare velocity of light $\nu_t$ is only tuned at 
the smaller volume $8^3 \cdot 32$ by demanding $c(0) = 1$, 
this value of $\nu_t$ is then adopted in
the simulation at the larger volume $16^3 \cdot 64$. 
Also note that $c(0)$ is extrapolated from the energies 
at momenta 0, $\frac{2\pi}{L}(1, 0, 0)$ and $\frac{2\pi}{L}(2, 0, 0)$. 
Since the spatial extension $L$ of the $16^3 \cdot 64$ run is twice 
that of the $8^3 \cdot 32$ run given the same lattice spacing, 
the extrapolation of $c(0)$ at the larger volume actually uses 
a different set of momenta.
Thus the consistency between the two $c(0)$'s also serves as 
an important check of the dispersion relation. 

The $16^3 \cdot 64$ run is used in the continuum extrapolation 
because its mass splittings are more accurate.

\subsection{Dispersion relation}

Compared with other lattice approaches to heavy quark systems, 
our approach has two distinctions:
1. the temporal lattice spacing is finer than the spatial lattice spacing;
2. the relativity of the lattice action 
   (broken in all heavy quark approaches)
   is restored numerically. 
Recall that we tune the bare velocity of light $\nu_t$ 
so that the $c(0)$ of the pseudo-scalar $1^1S_0$ state is close to 1. 
For the validation 
of our method of restoring relativity on an anisotropic lattice, 
it is important to check the universality of $c(0)$. 
We have checked it once in the study of finite volume effects 
by comparing two $c(0)$'s on $8^3 \cdot 32$ and $16^3 \cdot 64$ lattices
at $\beta = 5.7$. 
A more extensive check on the dispersion relation was done 
in the $\beta = 6.1$ run on a $16^3 \cdot 64$ lattice. 

The universality of $c(0)$ indeed holds (see table \ref{tab:dispersion})
for a variety of particles $1^1S_0$, $1^3S_1$ and $1^3P_0$ 
and holds for the extrapolations from two sets of momenta,
one set is $[0, \frac{2\pi}{L}(1, 0, 0), \frac{2\pi}{L}(2, 0, 0)]$, 
another set is $[0, \frac{2\pi}{L}(1, 1, 0), \frac{2\pi}{L}(2, 2, 0)]$.
Other particles and momenta are missing purely due to 
failures in data fitting.

Compared with the ``non-relativistic reinterpretation''
(to be described in this paragraph), our approach of 
maintaining the relativity of an anisotropic or heavy-quark Wilson action 
is conceptually clean.
Furthermore, it has a statistical advantage as well. 
It is argued \cite{Fermilab} that
without restoring relativity,  
the dispersion relation may be expanded in velocity as 
\beqs \label{eq:e_theirs}
E(\bp) & = & m_{\rm static} + 
             \frac{\bp^2}{2 m_{\rm kinetic}} + O(p^4) \\\nonumber
       & = & m_{\rm static} + \frac{\bp^2}{2 m_{\rm kinetic}(\bp)} \,
\eeqs
in which the kinetic mass $m_{\rm kinetic}$ is the true mass $m$, 
yet a difference between two static masses 
$\triangle m_{\rm static}$ also provides the true mass splitting
as does $\triangle m_{\rm kinetic}$.
The $m_{\rm kinetic}$ is obtained in the same way we use to obtain $c(0)$,
namely by extrapolation from $c(\bp)$ or $m_{\rm kinetic}(\bp)$
at the two lowest on-axis momenta. 
Combining the $c(\bp)$ equation
\beqs \label{eq:e_ours}
E(\bp)^2   & = &   m^2 + \bp^2 c(\bp)^2        
\eeqs
with the eq.(\ref{eq:e_theirs}) gives
\beq \label{eq:e_together}
  m_{\rm kinetic} = \frac{m_{\rm static}}{c(0)^2}. 
\eeq

As checked in \cite{KlassenQuark1,KlassenQuark2}, eq.(\ref{eq:e_together}) is 
indeed true within errors and 
the relative errors of $m_{\rm kinetic}$ are indeed roughly twice 
those of $c(0)$ as predicted by this equation. 
The absolute errors of $m_{\rm kinetic}$ are also found 
\cite{KlassenQuark1,KlassenQuark2} to be one order of magnitude larger than 
those of $m_{\rm static}$. 
This difference in statistical errors is not hard to understand: 
the kinetic mass comes out of a correlated fit 
on three meson correlators ({\it i.e.} from three momentum values), 
while the static mass is given by the meson correlator 
for zero momentum alone.
Presumably in the calculation of hadronic correlators, 
there is also extra difficulty in 
finding a smearing or box source size that is good for all three momenta.

\subsection{The charmonium spectrum}

In figure \ref{fig:charmonium} we plot 
the mass spectrum from quenched simulations on four anisotropic lattices 
against the experimental values \cite{Experiment}, 
and in table \ref{tab:charmonium} we list the precise numbers.
We can see that at this scale the agreement between our lattice simulations
and the observed values in Nature is very impressive and 
that even the effect of quenching is hard to see. 

The masses come out of the computer in units $\frac{1}{a_t}$, 
the inverse of the temporal lattice spacing. 
Since the lattice spacing is never an input and 
is always treated as 1 in simulations, 
in order to quote the masses in the physical unit GeV or MeV, 
we have to know the physical size of the lattice spacing $a_t$
(or equivalently that of $a_s$ 
as the true anisotropy $\xi$ has been fixed at 2).
In addition, the bare quark mass $m_0$ has to be tuned 
to correspond to the charm quark mass
so to obtain the charmonium mass spectrum. 
In determining the lattice scale $a$ we have not used any meson masses,
instead, 
we set the lattice scale using the Sommer scale $r_0$ because it can be 
measured more accurately than the popular choice of 
the $1^1P_1 - 1S$ mass splitting.
To fix the bare quark mass, 
we set the spin average $1S$ meson mass $m(1S)$ to its experimental value.
All the remaining energies, both the $n=1$ ground states and 
the $n=2$ excited states, are then predictions.

Among these predictions, most of the ground states 
deviate from experiment by less than 30 MeV. 
Regarding the minor discrepancies seen among the four estimates of 
each ground state, at least 50\% may be attributed to the 
initial choice of quark mass (which is tuned accurately to 1\%).
For example, there is a downward shift on 
all the masses from the $\beta=6.1$ run 
because the initial quark mass is slightly too small.
For the excited states 
the deviations from experiment are typically under 100 MeV.
Note there are no experimental data available for the excited states of 
particle $h_c$, $\chi_{c0}$ and $\chi_{c1}$.

The statistical errors on the excited states are one order of 
magnitude larger than the errors on the ground states for 
the following reasons.
The signal of an excited state, being proportional to 
exp$(-m_{\rm excited})$,
dies out much faster than the signal of a ground state, 
therefore far fewer time slices are useful in a fit determing
the excited state mass.
Furthermore, in our calculations 
the excited state signals are always mixed with
the larger ground state signals,
making the fitting of an excited state subject to 
the errors from the fitting of the ground state.
For reasons mentioned earlier, 
the errors listed in the data tables are 
purely statistical errors, thus
they do not include the systematic errors from the Sommer scale setting, 
the quark mass tuning, or from quenching.

\subsection{$1^1P_1 - 1S$ splitting}  \label{section:ps_splitting}
 
In lattice simulations of heavy quarks 
the $1^1P_1 - 1S$ splitting is often used to set the lattice scale, 
denoted as $a_{1^1P_1-1S}$.
Because the Sommer scale $r_0$ can be measured more accurately, 
We have used $r_0 = 0.5$ fm to set the scale, denoted as $a_{r_0}$.
In order to see how these two methods of scale setting differ, 
we plot our results (see figure \ref{fig:SP} and table \ref{tab:splitting})
in the form of their ratio 
$a_{1^1P_1-1S}/a_{r_0}$ 
\beq
\frac{a_{1^1P_1-1S}}{a_{r_0}} = 
       \frac{\triangle m_{1^1P_1 - 1S}}{458.5 {\rm  MeV}}
\eeq
where 
$\triangle m_{1^1P_1 - 1S}$ is the $1^1P_1 - 1S$ 
splitting with physical units from setting $r_0 = 0.5$ fm, and 
458.5 MeV is the experimental value for  the $1^1P_1 - 1S$ splitting.

The continuum $a^2$ extrapolation gives the ratio 
$a_{1^1P_1-1S}/a_{r_0} = 0.94(1)$. 
This discrepancy from 1 may come for two reasons.
One is due to quenching, 
the splitting $1^1P_1 - 1S$ is smaller than its physical value,
therefore $a_{1^1P_1-1S}$ has been underestimated.
Another reason is associated with $r_0 = 0.5$ fm, 
which is only a phenomenological estimate and 
not a hard experimental number.
Any errors with the assignment $r_0 = 0.5$ fm will affect our final 
values of masses quoted in physical units 
but only these final numbers.
We also show the most recent results 
based on the Fermilab approach \cite{Simone,Boyle} for comparison, 
where the agreements are obvious.

\subsection{$1^3S_1 - 1^1S_0$ splitting}
Here comes the most exciting part of our results: 
the hyper-fine structure of the charmonium mass spectrum, 
plotted in figure \ref{fig:SS} and listed in table \ref{tab:splitting}.
From the continuum $a^2$ extrapolation, 
the mass splitting $\triangle m_{1^3S_1 - 1^1S_0}$ 
comes out to be 71.8(20) MeV, 
which is 39\% smaller than its observed value of 117.1(2) MeV in nature.

For really heavy quarks, one may speculate that 
the hyper-fine splitting will be dominated by one-gluon exchange.
Thus one might be able to 
estimate the quenching effect by looking into 
the difference in the running of the strong coupling constant 
in quenched and full QCD.
The 39\% discrepancy from our simulations supports 
the expectation of large quenching effect on the hyper-fine splitting, 
although it is probably too aggressive for us to claim 
a 39\% quenching effect without qualification 
since we have not numerically tuned the clover term coefficients. 

As to the comparison with other lattice approaches, 
our results are consistent with calculations 
from the Fermilab approach \cite{Simone,Boyle}.
Also shown in figure \ref{fig:SS} are 
the NRQCD results \cite{NRQCD_SS1,NRQCD_SS2} at their best.
As realized and fully discussed in \cite{NRQCD_SS1,NRQCD_SS2},
the NRQCD results can not be trusted.
Inconsistent results are given by 
actions which differ only in the order of relativistic correction 
or which differ only in the tadpole prescription for quantum correction.
Therefore the NRQCD results will not be included in our later comparisons.

\subsection{$1^3P_1 - 1^3P_0$ splitting}

The $P$-wave fine structure of the charmonium mass spectrum is 
shown in figure \ref{fig:PP}, 
with exact numbers listed in table \ref{tab:splitting}.
The continuum extrapolation says the mass splitting of 
$1^3P_1 - 1^3P_0$ is 65(3) MeV, which is 30(5)\% smaller than 
the experimental value of 93(3) MeV.
As in the case of $S$-wave splitting, 
most of the discrepancy with experiment is 
attributed to the quenching effect
and it is not hard to claim consistency with 
results from the Fermilab approach \cite{Simone,Boyle}.

\subsection{Effects on results from small changes of bare parameters}

Here we discuss how the outputs
(masses, mass splittings and the relativity indicator $c(0)$) 
respond to
a 5\% or 10\% change of simulation inputs.
We will examine four inputs:
the bare quark mass $m_0$, 
the bare velocity of light $\nu_t$, 
and the two clover term coefficients $C_{\rm sw}^s$ and $C_{\rm sw}^t$. 
In our calculations we have tuned the inputs $\nu_t$ and $m_0$ 
numerically so that $c(0) = 1$ and $m(1S) = 3.0676 \,{\rm GeV}$. 
By contrast, inputs $C_{\rm sw}^s$ and $C_{\rm sw}^t$ are estimated 
(not tuned) from mean links in Landau gauge 
using tree-level tadpole improvement. 
To help the tuning of $\nu_t$ and $m_0$ and 
to estimate the effects from the absence of 
numerically determined $C_{\rm sw}^s$ and $C_{\rm sw}^t$, 
we need to know the  quantitative sensitivity of our results to these inputs.

Detailed in table \ref{tab:tuning} and identified as run 0 to 6, 
seven tests are done for this purpose. 
Run 0 is to be compared with others.
As to the remaining six, 
each of them differs from run 0 only in one input parameter.
Now let's look at the results in table \ref{tab:tuning} to 
see the effects of each input one by one.
We have listed three types of mass splittings in the table, 
namely the spin-spin splitting $\triangle m_{1^3S_1 - 1^1S_0}$,
the spin-orbital splitting $\triangle m_{1^3P_1 - 1^3P_0}$, 
and the $S-P$ splitting $\triangle m_{1^1P_1 - 1S}$. 
However we will only focus on the spin-spin splitting as 
the characteristics of the other two are either the same or hard
to tell given their relatively larger errors.

In runs 1 and 2, the bare quark mass $m_0$ is changed by $\pm 5\%$ 
from its best value of 0.51 as tuned and used in run 0. 
The comparison of these three runs shows that 
$m_0$ has no effect on $c(0)$ or on mass {\em splittings}. 
However, as expected, 
an increase of the bare quark mass $m_0$ gives a boost to 
masses of its bound states, $m(1S)$ listed as an example.
The errors from the tuning of $m_0$ therefore drop out 
of the discussions of mass splittings. 

In runs 3 and 4, the bare velocity of light $\nu_t$ is changed by $\pm 5\%$ 
from its tuned best value of 1.01. 
Comparing run 0, 3 and 4, we see that 
an increase of $\nu_t$ reduces the values of 
masses, mass splittings and $c(0)$.
This observation agrees with 
what is indicated through a field redefinition:
putting the bare relativity factor in its more conventional place, 
i.e. in front of the spatial derivative, 
\beq  \label{eq:transform}
  \bar{q}(x) \left[       
                m_0 + \nu_t   \!\not\!D_t   + \!\not\!{\bf D}
		\right] q(x) 
  \longrightarrow
  \bar{q}(x) \left[       
                \frac{m_0}{\nu_t} + \!\not\!D_t + 
                \frac{1}{\nu_t} \!\not\!{\bf D}
		\right] q(x)   \, 
\eeq
we see that effectively $\frac{1}{\nu_t}$ is 
the bare velocity of light and $\frac{m_0}{\nu_t}$ is the bare quark mass, 
therefore a change of $\nu_t$ has adverse effects on masses and $c(0)$.
As the mass splittings do not change noticeably with the bare quark mass
$m_0$ or  $\frac{m_0}{\nu_t}$, 
their dependence on $\nu_t$ has to be explained in some other way, 
which we do not yet know.

In runs 5 and 6, 
the two clover term coefficients $C_{\rm sw}^s$ and $C_{\rm sw}^t$ 
have been increased by 10\% one at a time 
over their estimated values used in run 0. 
The observation is that 
an increase of $C_{\rm sw}^s$ or $C_{\rm sw}^t$ 
has no noticeable effect on the value of $c(0)$,
but reduces meson masses, yet increases spin-spin splittings, 
while the effects from a 10\% increase of $C_{\rm sw}^s$ are 
much larger than that from a similar increase in $C_{\rm sw}^t$.
As the clover terms
enter the lattice action locally just like $m_0$, 
they are not expected to influence $c(0)$.
However, they are well-known in light quark calculations 
to make a positive contribution to meson masses
\footnote{In other words, the critical quark mass where 
pion becomes massless is less negative 
when the clover terms are added.}.
As to the hyperfine splitting $1^3S_1 - 1^1S_0$, 
which comes from the spin-spin interaction between two charm quarks, 
we first note that 
the spatial clover terms corresponds to lattice corrections to 
the chromo-magnetic coupling $\sigma \cdot \bf{B}$. 
Hence as is expected, the spin splitting is subject to 
the values of the $C_{\rm sw}^s$ clover coefficient.

\subsection{$\nu_s$-tuning vs. $\nu_t$-tuning} \label{section:tuning}

The above observation of the influence of $C_{\rm sw}^s$ and $C_{\rm sw}^t$ 
on the mass splittings is very important, 
especially since in our calculations
$C_{\rm sw}^s$ and $C_{\rm sw}^t$ are only (possibly very well) estimated.
While the quenching effects will remain, 
some percentages of the discrepancies between our results 
and the experimental data may simply disappear 
\footnote{Or the other way around,
which is less likely since in isotropic cases 
the numerical clover coefficients are found to be larger
than the tree-level estimates, thus the splittings will 
be increased toward experimental data.
}
when the numerical determination of 
$C_{\rm sw}^s$ and $C_{\rm sw}^t$ becomes feasible in the future. 
While this determination (say, by applying the Schr\"{o}dinger functional 
on an anisotropic lattice for heavy quarks) 
may be daunting both theoretically and computationally, 
this discussion of the sensitivity of our results to 
$C_{\rm sw}^s$ and $C_{\rm sw}^t$ 
leads us to comment on 
the choice of $\nu_t$-tuning over $\nu_s$-tuning in this work.

In the initial work \cite{KlassenQuark1,KlassenQuark2}, 
most calculations were done with the $\nu_s$-tuning.
If $C_{\rm sw}^s$ and $C_{\rm sw}^t$ are known numerically
for the values of $m_0$ and $\beta$ that enter into our simulations, 
\footnote{
Right now we ignore the $\xi$ dependence of $C_{\rm sw}^s$ and $C_{\rm sw}^t$ 
as we have fixed the value of the renormalized anisotropy $\xi$.
}
it should not matter which way is chosen to restore the relativity 
of the anisotropic lattice action, 
results from these two ways should agree 
even at finite lattice spacings up to $O(a^2)$ errors. 
If for both $\nu_t$-tuning and $\nu_s$-tuning, 
$C_{\rm sw}^s$ and $C_{\rm sw}^t$ have weak or linear dependency 
on $a_t m_0$ as we certainly have hoped for, 
it would still be legitimate for the $a^2$ extrapolation to apply,
so that at zero lattice spacing we would end up with the same results.
Unfortunately in the earlier work \cite{KlassenQuark1,KlassenQuark2},  
$\nu_t$-tuning and $\nu_s$-tuning 
were {\em not} found to give the same results in the continuum limit, 
at least not by using the clover coefficients estimated from 
eq.({\ref{eq:clover_term_estimate}).
Therefore at least in one of these two approaches, 
the dependence of $C_{\rm sw}^s$ and $C_{\rm sw}^t$ on $a_t m_0$ 
may be too strong to justify the $a^2$ extrapolation
for the values of $a$ studied here.

Based on the classical improvement 
discussed in section \ref{section:better}, 
we suspect that:
1. The two tunings would have been much more nearly consistent if 
   the tadpole improvement was {\em not} based on 
   the tree-level estimate given in 
   eq.(\ref{eq:nu_s_tuning})-(\ref{eq:nu_t_tuning}), 
   but instead came from the better classical estimate given by  
   eq.(\ref{eq:nu_s_tuning_better})-(\ref{eq:nu_t_tuning_better}).
2. If we estimate the clover coefficients by applying 
   tadpole corrections to the possibly less precise estimates in 
   eq.(\ref{eq:nu_s_tuning})-(\ref{eq:nu_t_tuning}), 
   which has been our practice so far, 
   the results from $\nu_t$-tuning should be more reliable than 
   the results from $\nu_s$-tuning.

Here is why. 
Without numerical determination of the clover coefficients, 
the crucial assumption or hope underlying the $a^2$ continuum 
extrapolation is that the clover coefficients 
depend on mass weakly or linearly.
Classically 
for the $\nu_s$-tuning both $C_{\rm sw}^t$ and $C_{\rm sw}^s$  depend 
on $m_c a$, while for the $\nu_t$-tuning 
only $C_{\rm sw}^t$ depends on $m_c a$ and there is no mass dependence in
$C_{\rm sw}^s = \nu_s = 1$.
Furthermore, looking at the field redefinition in eq.(\ref{eq:transform}), 
we find it contradictory in estimating
$C_{\rm sw}^s$ and $C_{\rm sw}^t$ to use 
eq.({\ref{eq:clover_term_estimate}) for both tunings.
While this has been the practice so far, 
the clover coefficients would be effectively larger 
for the $\nu_s$- than for the $\nu_t$-tuning if $\nu_t > 1$ 
(thus the mass splittings would be larger too) 
and the other way around if $\nu_t < 1$, 
which indeed is what is qualitatively found in 
\cite{KlassenQuark1,KlassenQuark2}. 
From what we see in the 10\% change test for the clover coefficients, 
the resulting discrepancy should be quite pronounced 
since $\nu_s$ or $\nu_t$ deviates from 1 by 1-12\%, 
the range described in table \ref{tab:basic}.

In short, we expect the choice of tuning to be a minor issue if 
the clover coefficients are estimated in the better way described above, 
and it would not be an issue if 
$C_{\rm sw}^s$ and $C_{\rm sw}^t$ were known numerically.
Most likely, on an isotropic lattice 
this problem of the mass dependence of $C_{\rm sw}^s$ and $C_{\rm sw}^t$ 
is only going to be more severe.
In retrospect, we should still first tune the bare velocity of light with 
the clover coefficients estimated from 
eq.(\ref{eq:nu_s_tuning_better})-(\ref{eq:nu_t_tuning}), 
but once we know the tuned values of $\nu_t$ or $\nu_s$, 
we should plug them into   
eq.(\ref{eq:nu_s_tuning_better})-(\ref{eq:nu_t_tuning_better})
to get a better estimate of clover coefficients, 
and then use the better estimate in the following tunings and real runs.

\section{Conclusion}

By running quenched simulations using an anisotropic action, 
we have been able to predict more reliable masses within the charmonium family 
than has been done in previous lattice calculations.  
The masses of both the radial $n=1$ ground state and 
the $n=2$ first excitation have been computed for the particles
$\eta_c$ ($^1S_0$), 
$J/\psi$ ($^3S_1$),
$h_c$ ($^1P_1$), 
$\chi_{c0}$ ($^3P_0$), 
and $\chi_{c1}$ ($^3P_1$). 
On a finer scale, 
from the continuum extrapolation we have the $S$-wave hyperfine splitting 
$\triangle m_{1^3S_1 - 1^1S_0}$ of 71.8(20) MeV, 
the $P$-wave fine structure
$\triangle m_{1^3P_1 - 1^3P_0}$ of 65(3) MeV,
and the $1P-1S$ splitting 	
$\triangle m_{1^1P_1 - 1S}$ of 431(3) MeV, 
which agrees with other lattice approaches 
\cite{Simone,Boyle}. 

Our work shows the intrinsic benefit of an anisotropic lattice 
where the temporal lattice spacing is finer than the spatial one. 
At relatively low computational cost,  
on an anisotropic lattice 
the signals of a hadron correlator are good on more time slices.
This is important to 
the calculations involving heavy quarks or excited states as they 
die out fast on current isotropic lattices.
The space-time exchange symmetry, 
broken both on a heavy-quark action 
and (only more explicitly) on an anisotropic lattice, 
has been restored by tuning the bare parameter $\nu_t$ based on 
the dispersion relation 
without resorting to the ``kinetic mass prescription''. 

While all errors given in data tables are statistical errors, 
the biggest errors in our results should be attributed to 
the systematic errors from quenching. 
Besides that, we have not numerically determined 
the two clover term coefficients $C_{\rm sw}^s$ and $C_{\rm sw}^t$, 
while the numerical tuning has been done for all other simulation inputs. 
It will require significant theoretical and computational effort 
to get rid of errors coming from these two sources, 
which is equally true in other lattice approaches to heavy quark systems.
A feasible project in the near future 
is to run more simulations at finer lattice spacings.
By doing so we will either have more support for current estimations of 
$C_{\rm sw}^s$ and $C_{\rm sw}^t$,  
or we will see the breaking of the $a^2$ extrapolation 
and thus be forced to pursue a fully numerical determination of
$C_{\rm sw}^s$ and $C_{\rm sw}^t$.
Either way progress will be made.

\section*{Acknowledgments}
The numerical calculations were done on the 400 Gflop QCDSP computer
\cite{Chen:1998cg} at Columbia University. This research was supported
in part by the DOE under grant \# DE-FG02-92ER40699.
Ping Chen would like to thank Prof. Norman H. Christ for 
being Ping's Ph.D. thesis advisor. 
She is grateful also to all other current and former Columbia QCDSP members 
including Prof. Robert D. Mawhinney, 
Dr. Dong Chen, 
Calin Cristian, 
George Fleming, 
Dr. Chulwoo Jung,  
Dr. Adrian Kaehler, 
Xiaodong Liao, 
Guofeng Liu, 
Dr. Yubing Luo, 
Dr. Catalin Malureanu, 
Dr. Thomas Manke, 
Chengzhong Sui, 
Dr. Pavlos Vranas, 
Lingling Wu and 
Yuri Zhestkov. 
At last, this work would not have been here without the
significant contribution from Dr. Tim Klassen.

\bibliography{paper2000}

\begin{thebibliography}{10}
\expandafter\ifx\csname bibnamefont\endcsname\relax
  \def\bibnamefont#1{#1}\fi
\expandafter\ifx\csname bibfnamefont\endcsname\relax
  \def\bibfnamefont#1{#1}\fi
\expandafter\ifx\csname url\endcsname\relax
  \def\url#1{\texttt{#1}}\fi
\expandafter\ifx\csname urlprefix\endcsname\relax\def\urlprefix{URL }\fi
\expandafter\ifx\csname bibinfo\endcsname\relax \def\bibinfo#1#2{#2}\fi
\expandafter\ifx\csname eprint\endcsname\relax \def\eprint#1{#1}\fi

\bibitem{Lepage}
\bibinfo{author}{\bibfnamefont{G.~P.} \bibnamefont{Lepage}} \emph{et~al.},
  \bibinfo{journal}{Phys. Rev.} \textbf{\bibinfo{volume}{D46}},
  \bibinfo{pages}{4052} (\bibinfo{year}{1992}).

\bibitem{Davies}
\bibinfo{author}{\bibfnamefont{C.}~\bibnamefont{Davies}}
  \eprint{hep-ph/9710394}.

\bibitem{NRQCD_SS1}
\bibinfo{author}{\bibfnamefont{H.~D.} \bibnamefont{Trottier}},
  \bibinfo{journal}{Phys. Rev.} \textbf{\bibinfo{volume}{D55}},
  \bibinfo{pages}{6844} (\bibinfo{year}{1997}).

\bibitem{NRQCD_SS2}
\bibinfo{author}{\bibfnamefont{N.~H.} \bibnamefont{Shakespeare}}
  \bibnamefont{and} \bibinfo{author}{\bibfnamefont{H.~D.}
  \bibnamefont{Trottier}} \eprint{hep-lat/9802038}.

\bibitem{Fermilab}
\bibinfo{author}{\bibfnamefont{A.~X.} \bibnamefont{El-Khadra}},
  \bibinfo{author}{\bibfnamefont{A.~S.} \bibnamefont{Kronfeld}},
  \bibnamefont{and} \bibinfo{author}{\bibfnamefont{P.~B.}
  \bibnamefont{Mackenzie}}, \bibinfo{journal}{Phys. Rev.}
  \textbf{\bibinfo{volume}{D55}}, \bibinfo{pages}{3933} (\bibinfo{year}{1997}),
  \eprint{hep-lat/9604004}.

\bibitem{KlassenQuark1}
\bibinfo{author}{\bibfnamefont{T.~R.} \bibnamefont{Klassen}},
  \bibinfo{journal}{Nucl. Phys. B Proc. Suppl.} \textbf{\bibinfo{volume}{73}},
  \bibinfo{pages}{918} (\bibinfo{year}{1999}), \eprint{hep-lat/9809174}.

\bibitem{KlassenQuark2}
\bibinfo{author}{\bibfnamefont{T.~R.} \bibnamefont{Klassen}}
  \bibinfo{note}{Heavy quarks on anisotropic lattices (unpublished)}.

\bibitem{KlassenGauge}
\bibinfo{author}{\bibfnamefont{T.~R.} \bibnamefont{Klassen}},
  \bibinfo{journal}{Nucl. Phys.} \textbf{\bibinfo{volume}{B533}},
  \bibinfo{pages}{557} (\bibinfo{year}{1998}), \eprint{hep-lat/9803010}.

\bibitem{Creutz}
\bibinfo{author}{\bibfnamefont{M.}~\bibnamefont{Creutz}},
  \bibinfo{journal}{Phys. Rev.} \textbf{\bibinfo{volume}{D21}},
  \bibinfo{pages}{2308} (\bibinfo{year}{1980}).

\bibitem{Kennedy}
\bibinfo{author}{\bibfnamefont{A.~D.} \bibnamefont{Kennedy}} \bibnamefont{and}
  \bibinfo{author}{\bibfnamefont{B.~J.} \bibnamefont{Pendleton}},
  \bibinfo{journal}{Phys. Lett.} \textbf{\bibinfo{volume}{156B}},
  \bibinfo{pages}{393} (\bibinfo{year}{1985}).

\bibitem{Cabibbo1}
\bibinfo{author}{\bibfnamefont{N.}~\bibnamefont{Cabibbo}} \bibnamefont{and}
  \bibinfo{author}{\bibfnamefont{E.}~\bibnamefont{Marinari}},
  \bibinfo{journal}{Phys. Lett.} \textbf{\bibinfo{volume}{119B}},
  \bibinfo{pages}{387} (\bibinfo{year}{1982}).

\bibitem{KlassenScale1}
\bibinfo{author}{\bibfnamefont{R.}~\bibnamefont{Sommer}},
  \bibinfo{journal}{Nucl. Phys.} \textbf{\bibinfo{volume}{B411}},
  \bibinfo{pages}{839} (\bibinfo{year}{1994}), \eprint{hep-lat/9310022}.

\bibitem{KlassenScale2}
\bibinfo{author}{\bibfnamefont{R.~G.} \bibnamefont{Edwards}},
  \bibinfo{author}{\bibfnamefont{U.~M.} \bibnamefont{Heller}},
  \bibnamefont{and} \bibinfo{author}{\bibfnamefont{T.~R.}
  \bibnamefont{Klassen}}, \bibinfo{journal}{Nucl. Phys.}
  \textbf{\bibinfo{volume}{B517}}, \bibinfo{pages}{377} (\bibinfo{year}{1998}),
  \eprint{hep-lat/9711003}.

\bibitem{KlassenScale3}
\bibinfo{author}{\bibfnamefont{R.~G.} \bibnamefont{Edwards}},
  \bibinfo{author}{\bibfnamefont{U.~M.} \bibnamefont{Heller}},
  \bibnamefont{and} \bibinfo{author}{\bibfnamefont{T.~R.}
  \bibnamefont{Klassen}} \bibinfo{note}{Unpublished}.

\bibitem{CG}
\bibinfo{author}{\bibfnamefont{M.~R.} \bibnamefont{Hestenes}} \bibnamefont{and}
  \bibinfo{author}{\bibfnamefont{E.}~\bibnamefont{Stiefel}},
  \bibinfo{journal}{Journal of Research of the National Bureau of Standards}
  \bibinfo{note}{Vol. 49, No.6, December 1952 Research Paper 2379}.

\bibitem{Recipe}
\bibinfo{author}{\bibfnamefont{W.}~\bibnamefont{Press}},
  \bibinfo{author}{\bibfnamefont{B.}~\bibnamefont{Flannery}},
  \bibinfo{author}{\bibfnamefont{S.}~\bibnamefont{Teukolsky}},
  \bibnamefont{and}
  \bibinfo{author}{\bibfnamefont{W.}~\bibnamefont{Vetterling}}
  \bibinfo{note}{Numerical Recipes in C: The Art of Scientific Computing}.

\bibitem{Experiment}
\bibinfo{author}{\bibfnamefont{C.}~\bibnamefont{Caso}} \emph{et~al.}
  (\bibinfo{collaboration}{Particle Data Group}), \bibinfo{journal}{European
  Physical Journal} \textbf{\bibinfo{volume}{C3}}, \bibinfo{pages}{1}
  (\bibinfo{year}{1998}).

\bibitem{Simone}
\bibinfo{author}{\bibfnamefont{J.}~\bibnamefont{Simone}} \bibinfo{note}{Private
  communication on Fermilab results}.

\bibitem{Boyle}
\bibinfo{author}{\bibfnamefont{P.}~\bibnamefont{Boyle}}
  (\bibinfo{collaboration}{UKQCD collaboration}) \eprint{hep-lat/9903017}.

\bibitem{Chen:1998cg}
\bibinfo{author}{\bibfnamefont{D.}~\bibnamefont{Chen}} \emph{et~al.},
  \bibinfo{journal}{Nucl. Phys. Proc. Suppl.} \textbf{\bibinfo{volume}{73}},
  \bibinfo{pages}{898} (\bibinfo{year}{1999}), \eprint{hep-lat/9810004}.

\end{thebibliography}


\tabcolsep0.4cm


\begin{table}[ptb]
\begin{center}
\begin{tabular}{lccccc} 
$\beta$  	
		& 5.6			& 5.7
		& 5.7
		& 5.9			& 6.1
								\\
$L^3 \cdot T$	
		& $8^3 \cdot 32$	& $8^3 \cdot 32$ 
		& $16^3 \cdot 64$    	
		& $16^3 \cdot 64$	& $16^3 \cdot 64$    	
								\\
$\xi_0$ ($\xi=2$)
		& 1.632156		& 1.654729
		& 1.654729
		& 1.690713              & 1.718306          	
								\\
$m_0 \cdot a_t$  	
		& 0.69 			& 0.51  & 0.51  
		& 0.195              	& 0.05
								\\
$C_{sw}^s$ 	
		& 2.364			& 2.138  & 2.138
		& 1.889			& 1.7614
								\\
$C_{sw}^t$ 	
		& 1.429			& 1.3252 & 1.3252
		& 1.2055		& 1.1431
								\\
$\nu_s$ 	
		& 1 & 1 & 1 & 1 & 1
								\\
$\nu_t$ 	
		& 0.92			& 1.01 & 1.01
		& 1.09			& 1.12
								\\\hline
$\langle U_s^{\rm Landau Gauge} \rangle$ 
		& 0.7504(2)		& 0.7762(2)  & 0.7762(2)
		& 0.8091(2)		& 0.8280(2)
								\\
$\langle U_t^{\rm Landau Gauge} \rangle$ 
		& 0.9321(1)		& 0.9394(1) & 0.9394(1)
		& 0.9504(1)		& 0.9569(1)
								\\\hline
$c(0)$
		& 1.012(2)      	& 1.000(2)
		& 0.991(3)				
		& 0.984(3)  		& 0.984(3)  
								\\
$m(1S)_{\rm lat}$ (GeV)
		& 3.063(1)		
		& 3.079(2)		& 3.078(2)
		& 3.069(2)    		& 3.044(2)
								\\
$m(1S)_{\rm exp}$ (GeV)
		& 3.0676(1)  & 3.0676(1)  & 3.0676(1)  & 3.0676(1)
		& 3.0676(1)
								\\\hline
configurations
		& 1480			& 1350
		& 440				
		& 410			& 588
								\\
spatial L (fm)
		& 2.02			& 1.66
		& 3.32
		& 2.17			& 1.54
								\\
\end{tabular}
\end{center}
\caption[Basic simulation parameters]{
This table consists of four parts.
From top to bottom: 
1. basic simulation inputs;  
2. mean links measured in Landau gauge
 used to estimate the clover coefficients;
3. tuning accuracy of the effective velocity of light $c(0)$ 
  and the 1S spin average meson mass;
4. number of measurements and the physical spatial extension.
All the errors on masses, $m(1S)$ in this table 
and more masses in the following
tables, are purely statistical errors.
}
\label{tab:basic}
\end{table}


\tabcolsep0.6cm


\begin{table}[ptb]
\begin{center}
\begin{tabular}{lcccc}
$\beta$  	
		& 5.6			& 5.7 	
		& 5.9			& 6.1
								\\
$\xi$
		& 2			& 2
		& 2			& 2
								\\\hline
$\xi_0$
		& 1.632156		& 1.654729
		& 1.690713              & 1.718306          	
								\\
$r_{0N}/a_s$ 
		& 1.982(10) 		& 2.413(6)
		& 3.690(11)             & 5.207(29)
								\\\hline
L (fm) if $\frac{L}{a_s}=8$
		& 2.018			& 1.658
		& 1.084                 & 0.768
								\\
L (fm) if $\frac{L}{a_s}=16$
		& 4.036			& 3.315
		& 2.168                 & 1.536
								\\
$1/a_t$ in GeV
		& 1.564			& 1.905
		& 2.913			& 4.110
								\\
\end{tabular}
\end{center}
\caption[Scale setting]{ 
At a given $\beta$ and true anisotropy $\xi$,
the corresponding bare anisotropy $\xi_0$ and 
the Sommer scale $r_{0N}/a_s$ are 
quoted from \cite{KlassenGauge,KlassenScale2,KlassenScale3}.
Both were determined to 1\% accuracy. 
On the last line $1/a_t$ is given in GeV, 
numbers which are used in this work to express 
the charmonium mass spectrum in physical units.
}
\label{tab:scale}
\end{table}


\tabcolsep0.8cm


\begin{table}[ptb]
\begin{center}
\begin{tabular}{ccccc} 
	  $\Gamma$			& $^{2S+1}L_J$
	& $J^{PC}$			
	& $u\bar{u}$			& $c\bar{c}$
								\\\hline
	  $\gamma_5$			& $^1S_0$
	& $0^{-+}$			
	& $\pi$				& $\eta_c$
								\\
	  $\gamma_s$			& $^3S_1$
	& $1^{--}$			
	& $\rho$			& $J/\psi$
								\\
	  $\gamma_s\gamma_{s\prime}$	& $^1P_1$
	& $1^{+-}$			
	& $b_1$				& $h_c$
								\\
	  $1$				& $^3P_0$
	& $0^{++}$			
	& $a_0$				& $\chi_{c0}$
								\\
	  $\gamma_5\gamma_{s}$		& $^3P_1$
	& $1^{++}$			
	& $a_1$				& $\chi_{c1}$
								\\
\end{tabular}
\end{center}
\caption[Meson operators]{ 
Meson states created by 
local and relativistic operators of the form $\bar{\psi}\Gamma\psi$, 
labelled in spectroscopic notation and by the particles 
they are identified with in the system of light flavors $u\bar{u}$ 
and the charmonium family $c\bar{c}$.
}
\label{tab:meson_op}
\end{table}


\tabcolsep0.1cm


\begin{table}[ptb]
\begin{center}
\begin{tabular}{lccccc} 
$\beta$
		& 5.6			& 5.7   
		& 5.7   
		& 5.9			& 6.1
								\\
$L^3 \cdot T$
		& $8^3 \cdot 32$	& $8^3 \cdot 32$ 
		& $16^3 \cdot 64$    	
		& $16^3 \cdot 64$	& $16^3 \cdot 64$    	
								\\\hline
Therm sweeps
		& 2500 			& 2500  & 2500  
		& 5000			& 5000
								\\\hline
Landau gauge fixing
		& \multicolumn{5}{c}
		{all elements of the traceless antihermitian 
		part $\leq 10^{-5}$}	
								\\
updating 
		& 100   & 100   & 100
		& 500			& 500
								\\
Landau link config 
		& 80		 & 15 & 15
		& 10			& 10
								\\\hline
Spectrum updating 
		& 100			& 100	& 100
		& 200 			& 400
								\\
sink \& source
		& \multicolumn{5}{c}
		{local sink, three box sources of different sizes}	
								\\
source $(N_x \cdot N_y \cdot N_z)_s$ 
		& $2\cdot2\cdot2$	& $2\cdot2\cdot2$
		& $2\cdot2\cdot2$
		& $3\cdot3\cdot4$	& $4\cdot4\cdot4$	
								\\
source $(N_x \cdot N_y \cdot N_z)_m$ 
		& $3\cdot3\cdot3$	& $4\cdot4\cdot4$
		& $4\cdot4\cdot4$
		& $5\cdot6\cdot6$	& $7\cdot8\cdot8$	
								\\
source $(N_x \cdot N_y \cdot N_z)_l$ 
		& $6\cdot6\cdot6$	& $6\cdot6\cdot6$
		& $6\cdot6\cdot6$
		& $11\cdot11\cdot11$	& $13\cdot13\cdot13$	
								\\
Coulomb gauge fixing 
		& \multicolumn{5}{c}
		{sum of square of all 18 matrix elements  $\leq 10^{-9}$}
								\\\hline
CG stopping cnd
		& 1.0e-8
		& 1.0e-8
		& 1.0e-8
		& 1.0e-8
		& 1.0e-8
								\\
CG iterations
		& 42(4)			& 43(3)
		& 49(4)			
		& 64(4)			& 85(5)
								\\\hline
Mean plaq $\langle \Box \rangle_s$
		& 0.41318(7)		& 0.43925(6)
		& 0.43918(3)
		& 0.47723(2)		& 0.50295(2)
								\\
Mean plaq $\langle \Box \rangle_t$
		& 0.65693(3)		& 0.67401(3)
		& 0.67398(1)
		& 0.69923(1)		& 0.71722(1)
								\\
\end{tabular}
\end{center}
\caption[Measurement conditions]{ 
Simulation parameters detailed in six parts from top to bottom: 
1. inputs to identify each simulation;
2. heatbath sweeps for thermalization;
3. Landau gauge fixing condition, 
   heatbath updating sweeps, 
   and number of measurements of mean links;
4. heatbath updating sweeps, 
   sink and source types, 
   small-size box source, 
   medium-size box source,
   large-size box source, 
   and Coulomb gauge fixing condition for the mass spectrum;
5. stopping condition and number of iterations of 
   the conjugate gradient (CG) in the calculation of quark propagators;
6. measured average values of spatial and temporal plaquettes. 
}
\label{tab:sim_cnd}
\end{table}


\tabcolsep0.2cm


\begin{table}[ptb]
\begin{center}
\begin{tabular}{lllllll} 
$\beta$  	
		& \multicolumn{1}{c}{5.6}			
		& \multicolumn{2}{c}{5.7} 	
		& \multicolumn{1}{c}{5.9}
		& \multicolumn{1}{c}{6.1}
		& Nature
								\\
$L^3 \cdot T$	
		& $8^3 \cdot 32$	& $8^3 \cdot 32$ 
		& $16^3 \cdot 64$    	
		& $16^3 \cdot64$	& $16^3 \cdot 64$    	
		& N/A	
								\\\hline
$m_{1^1S_0}$ (GeV)   	
		& 3.001(1) 			
		& 3.023(2)			& 3.021(2)
		& 3.012(1)
		& 2.992(2)
		& 2.9798(2)
								\\
$m_{2^1S_0}$ (GeV)  	
		& 3.61(3) 			
		& 3.70(6)			& 3.65(3)	
		& 3.70(2)
		& 3.66(3)
		& 3.594(5)
								\\
$m_{1^3S_1}$ (GeV)  		
		& 3.084(1) 		
		& 3.099(2)			& 3.098(2)		
		& 3.090(1)
		& 3.062(2)
		& 3.09688(4)
 								\\
$m_{2^3S_1}$ (GeV)  		 	
		& 3.65(3) 			
		& 3.64(3)			& 3.78(2)	
		& 3.75(4)
		& 3.73(3)
		& 3.68600(9)
								\\
$m_{1^1P_1}$ (GeV)  		 	
		& 3.523(3) 			
		& 3.526(5)			& 3.518(5)
		& 3.517(5)
		& 3.498(12)
		& 3.5261(2)
								\\
$m_{2^1P_1}$ (GeV)  		 	
		& 4.12(5) 
		& 3.92(10)			& 4.02(5)	
		& 4.09(5)
		& 3.86(10)
		& N/A
								\\
$m_{1^3P_0}$ (GeV)  		 	  
		& 3.499(2) 
		& 3.481(21)			& 3.480(2)		
		& 3.465(3)
		& 3.421(4)
		& 3.417(3)
								\\
$m_{2^3P_0}$ (GeV)  		 	  	
		& 3.82(5) 			
		& 3.79(8)			& 4.05(5)
		& 4.03(4)
		& 4.02(5)
		& N/A
								\\
$m_{1^3P_1}$ (GeV)  		 	  	
		& 3.530(2) 		
		& 3.523(6)			& 3.518(4)
		& 3.519(2)
		& 3.472(6)
		& 3.5105(1)
								\\
$m_{2^3P_1}$ (GeV)  		 	  	
		& 3.92(5) 			
		& 3.85(11)			& 4.08(6)
		& 3.99(5)
		& 4.00(8)
		& N/A 
								\\
\end{tabular}
\end{center}
\caption[Charmonium spectrum]{ 
Charmonium spectrum measured at four values of $\beta$, compared with
their observed values in nature.
As explained in the text, 
the errors from scale setting are {\em not} included,
which also applies to the tables that follow.
}
\label{tab:charmonium}
\end{table}



\begin{table}[ptb]
\begin{center}
\begin{tabular}{lccccccc} 
$\beta$  	
		& \multicolumn{1}{c}{5.6}			
		& \multicolumn{2}{c}{5.7} 	
		& \multicolumn{1}{c}{5.9}
		& \multicolumn{1}{c}{6.1}
		& $a \rightarrow 0 $	& Nature
								\\
$L^3 \cdot T$	
		& $8^3 \cdot 32$	& $8^3 \cdot 32$ 
		& $16^3 \cdot 64$    	
		& $16^3 \cdot64$	& $16^3 \cdot 64$    	
		& N/A			& N/A
								\\\hline
$\triangle m_{1^3S_1 - 1^1S_0}$ (MeV) 		 	  	
		& 84.7(4) 			
		& 75.9(5)			& 77.2(6)
		& 76.4(6)
		& 73.3(6)			
		& 71.8(20)
		& 117.1(2)
								\\
$\triangle m_{1^3P_1 - 1^3P_0}$ (MeV) 	 		 	  	
		& 33(1) 			
		& 37(3)				& 43(3)		
		& 59(1)
		& 58(2)			
		& 65(3)
		& 93(3)
								\\
$\triangle m_{1^1P_1 - 1S}$ (MeV) 		 	  	
		& 473(7) 			
		& 449(6)			& 454(6)
		& 442(5)
		& 439(5)
		& 431(3)
		& 458.5(2)
								\\
$\triangle m_{2^1S_0 - 1^1S_0}$ (MeV) 		 	  	  	
		& 611(28) 			
		& 675(57) 			& 625(27)
		& 691(20)
		& 666(34)
		& 693(19)
		& 614(5)
								\\
$\triangle m_{2^3S_1 - 1^3S_1}$ (MeV)   		 	
		& 563(32) 			
		& 546(29)           		& 677(20)
		& 662(35)
		& 665(34)
		& 696(40)
		& 589.1(1)
								\\
$\triangle m_{2^1P_1 - 1^1P_1}$ (MeV)  		 	
		& 591(48) 			
		& 391(103)        		& 503(48)
		& 575(52)
		& 358(102)
		& 412(92)
		& N/A
								\\
$\triangle m_{2^3P_0 - 1^3P_0}$ (MeV) 	 		 	  	
		& 321(52) 			
		& 307(76)			& 567(50)
		& 566(41)
		& 603(53)
		& 667(75)
		& N/A
								\\
$\triangle m_{2^3P_1 - 1^3P_1}$  (MeV) 	 		 	  	
		& 385(52)
		& 327(105)       		& 559(55)
		& 476(53)
		& 532(78)
		& 549(71)        
		& N/A 
								\\
\end{tabular}
\end{center}
\caption[Continuum extrapolation of the mass differences]{ 
Mass splittings in the charmonium spectrum measured at four values of $\beta$, 
their extrapolated values to the continuum limit,
and the observed values in Nature.
}
\label{tab:splitting}
\end{table}


\tabcolsep0.1cm


\begin{table}[ptb]
\begin{center}
\begin{tabular}{lccccccc} 
Fitted 
		& fitting			& $\chi^2$ per
		& Q				& goodness
		& dropped			& d.o.f.
		& bin				
								\\ 
quantities
		& range				& d.o.f.
		& value				& of fit
		& eigenvalues			& left 
		& size
								\\\hline
$c(\bp = 0)$  		 	  	
		& 6-13 			& 2.8(5)
		& 0.003			& 0.029
		& 9
		& 9 			& 2
								\\
$m_{1S}$, $\triangle m_{1^3S_1 - 1^1S_0}$, $\triangle m_{1^1P_1 - 1S}$
		& 6-13 			& 1.6(5)
		& 0.11			& 1.1
		& 8			
		& 10 			& 2
								\\
$\triangle m_{1^3P_1 - 1^3P_0}$  		 	  	
		& 7-13 			& 0.7(5)
		& 0.73			& 7.3
		& 0
		& 10 			& 2
								\\
$m_{1^1S_0}$, $\triangle m_{2^1S_0 - 1^1S_0}$  	
		& 4-13 			& 0.8(4)
		& 0.68			& 10.1			
		& 8 			& 15
		& 2 
								\\
$m_{1^3S_1}$, $\triangle m_{2^3S_1 - 1^3S_1}$  		 	
		& 5-13 			& 1.3(4)
		& 0.22			& 3.1
		& 6 			& 14
		& 2
								\\
$m_{1^1P_1}$, $\triangle m_{2^1P_1 - 1^1P_1}$  		 	
		& 2-13 			& 1.5(3)
		& 0.06			& 1.2
		& 8			& 21
		& 2
								\\
$m_{1^3P_0}$, $\triangle m_{2^3P_0 - 1^3P_0}$  		 	  	
		& 3-13 			& 1.0(3)
		& 0.42			& 9.5
		& 3			& 23			
		& 2
								\\
$m_{1^3P_1}$, $\triangle m_{2^3P_1 - 1^3P_1}$  		 	  	
		& 3-13 			& 1.2(3)
		& 0.26			& 6.0
		& 3			& 23			
		& 2
								\\
\end{tabular}
\end{center}
\caption[Fitting details at $\beta = 5.6$ on an $8^3 \cdot 32$ lattice]{ 
Fitting details at $\beta = 5.6$ on an $8^3 \cdot 32$ lattice.
See the text for details on the fitting ansatz, procedure and criteria.
}
\label{tab:fit_b5.6}
\end{table}



\begin{table}[ptb]
\begin{center}
\begin{tabular}{lccccccc} 
Fitted 
		& fitting			& $\chi^2$ per
		& Q				& goodness
		& dropped			& d.o.f.
		& bin				
								\\ 
quantities
		& range				& d.o.f.
		& value				& of fit
		& eigenvalues			& left 
		& size
								\\\hline
$c(\bp = 0)$  		 	  	
		& 8-15 			& 0.9(4)
		& 0.58			& 7.5
		& 5
		& 13 			& 2
								\\
$m_{1S}$, $\triangle m_{1^3S_1 - 1^1S_0}$, $\triangle m_{1^1P_1 - 1S}$
		& 7-15 			& 1.3(4)
		& 0.22			& 3.1
		& 7
		& 14 			& 2
								\\
$\triangle m_{1^3P_1 - 1^3P_0}$ 
		& 7-15 			& 1.2(4)
		& 0.24			& 3.4
		& 0
		& 14 			& 2
								\\
$m_{1^1S_0}$, $\triangle m_{2^1S_0 - 1^1S_0}$  
		& 8-15 			& 1.5(4)
		& 0.13			& 1.6
		& 5 			& 12
		& 2 
								\\
$m_{1^3S_1}$, $\triangle m_{2^3S_1 - 1^3S_1}$ 
		& 8-15 			& 1.2(4)
		& 0.26			& 3.8
		& 2 			& 15
		& 2
								\\
$m_{1^1P_1}$, $\triangle m_{2^1P_1 - 1^1P_1}$ 
		& 5-15 			& 1.0(3)
		& 0.44			& 8.8
		& 6			& 20
		& 2
								\\
$m_{1^3P_0}$, $\triangle m_{2^3P_0 - 1^3P_0}$ 
		& 5-15 			& 1.6(3)
		& 0.04			& 0.9
		& 5			& 21
		& 2
								\\
$m_{1^3P_1}$, $\triangle m_{2^3P_1 - 1^3P_1}$ 
		& 5-15 			& 0.9(3)
		& 0.63			& 13
		& 6			& 20
		& 2
								\\
\end{tabular}
\end{center}
\caption[Fitting details at $\beta = 5.7$ on an $8^3 \cdot 32$ lattice]{ 
Fitting details at $\beta = 5.7$ on an $8^3 \cdot 32$ lattice.
}
\label{tab:fit_b5.7_L8}
\end{table}



\begin{table}[ptb]
\begin{center}
\begin{tabular}{lccccccc} 
Fitted 
		& fitting			& $\chi^2$ per
		& Q				& goodness
		& dropped			& d.o.f.
		& bin				
								\\ 
quantities
		& range				& d.o.f.
		& value				& of fit
		& eigenvalues			& left 
		& size
								\\\hline
$c(\bp = 0)$  		 	  	
		& 6-16  		& 2.5(3)
		& 0.001			& 0.02
		& 13
		& 14 			& 2
								\\
$m_{1S}$, $\triangle m_{1^3S_1 - 1^1S_0}$, $\triangle m_{1^1P_1 - 1S}$
		& 5-15 			& 1.4(3)
		& 0.14			& 2.5			
		& 9 			& 18
		& 2 
								\\
$\triangle m_{1^3P_1 - 1^3P_0}$  	
		& 5-15 			& 0.7(4)
		& 0.85			& 14
		& 2
		& 16 			& 2
								\\
$m_{1^1S_0}$, $\triangle m_{2^1S_0 - 1^1S_0}$ 
		& 4-15 			& 1.0(3)
		& 0.42			& 8.0
		& 10			
		& 19 			& 2
								\\
$m_{1^3S_1}$, $\triangle m_{2^3S_1 - 1^3S_1}$ 
		& 3-15 			& 1.1(3)
		& 0.38			& 7.2		
		& 13 			
		& 19			& 2
								\\
$m_{1^1P_1}$, $\triangle m_{2^1P_1 - 1^1P_1}$ 
		& 3-15 			& 1.2(3)
		& 0.19			& 4.7
		& 8			& 24
		& 2
								\\
$m_{1^3P_0}$, $\triangle m_{2^3P_0 - 1^3P_0}$  
		& 3-15 			& 1.2(2)
		& 0.19			& 5.9
		& 0			& 32			
		& 2
								\\
$m_{1^3P_1}$, $\triangle m_{2^3P_1 - 1^3P_1}$
		& 3-15 			& 1.1(3)
		& 0.31			& 7.5
		& 8			& 24			
		& 2
								\\
\end{tabular}
\end{center}
\caption[Fitting details at $\beta = 5.7$ on a $16^3 \cdot 64$ lattice]{ 
Fitting details at $\beta = 5.7$ on a $16^3 \cdot 64$ lattice.
}
\label{tab:fit_b5.7_L16}
\end{table}



\begin{table}[ptb]
\begin{center}
\begin{tabular}{lccccccc} 
Fitted 
		& fitting			& $\chi^2$ per
		& Q				& goodness
		& dropped			& d.o.f.
		& bin				
								\\ 
quantities
		& range				& d.o.f.
		& value				& of fit
		& eigenvalues			& left 
		& size
								\\\hline
$c(\bp = 0)$  		 	  	
		& 9-24 			& 1.8(3)
		& 0.009			& 0.21
		& 19
		& 23 			& 2
								\\
$m_{1S}$, $\triangle m_{1^3S_1 - 1^1S_0}$, $\triangle m_{1^1P_1 - 1S}$
		& 8-24 			& 1.2(3)
		& 0.17			& 4.9
		& 16			
		& 29 			& 2
								\\
$\triangle m_{1^3P_1 - 1^3P_0}$ 
		& 5-24 			& 1.3(2)
		& 0.12			& 4.4
		& 0
		& 36 			& 2
								\\
$m_{1^1S_0}$, $\triangle m_{2^1S_0 - 1^1S_0}$ 
		& 6-24 			& 1.2(2)
		& 0.18			& 5.8
		& 17 			& 33 
		& 2 
								\\
$m_{1^3S_1}$, $\triangle m_{2^3S_1 - 1^3S_1}$ 
		& 7-24 			& 1.5(2)
		& 0.04			& 1.2			
		& 15 			& 32
		& 2
								\\
$m_{1^1P_1}$, $\triangle m_{2^1P_1 - 1^1P_1}$  
		& 5-24 			& 1.2(2)
		& 0.22			& 8.2
		& 15			& 38
		& 2
								\\
$m_{1^3P_0}$, $\triangle m_{2^3P_0 - 1^3P_0}$
		& 5-24 			& 1.4(2)
		& 0.05			& 2.3
		& 11			& 42
		& 2
								\\
$m_{1^3P_1}$, $\triangle m_{2^3P_1 - 1^3P_1}$
		& 6-24 			& 1.0(2)
		& 0.54			& 23
		& 7			& 43			
		& 2
								\\
\end{tabular}
\end{center}
\caption[Fitting details at $\beta = 5.9$ on a $16^3 \cdot 64$ lattice]{ 
Fitting details at $\beta = 5.9$ on a $16^3 \cdot 64$ lattice.
}
\label{tab:fit_b5.9}
\end{table}



\begin{table}[ptb]
\begin{center}
\begin{tabular}{lccccccc} 
Fitted 
		& fitting			& $\chi^2$ per
		& Q				& goodness
		& dropped			& d.o.f.
		& bin				
								\\ 
quantities
		& range				& d.o.f.
		& value				& of fit
		& eigenvalues			& left 
		& size
								\\\hline
$c(\bp = 0)$  		 	  	
		& 17-32 			& 1.3(3)
		& 0.15				& 4.1
		& 14
		& 28	 			& 2
								\\
$m_{1S}$, $\triangle m_{1^3S_1 - 1^1S_0}$, $\triangle m_{1^1P_1 - 1S}$
		& 13-32 		& 1.2(2)
		& 0.16			& 5.8
		& 17			
		& 37 			& 2
								\\
$\triangle m_{1^3P_1 - 1^3P_0}$ 
		& 12-32 		& 1.2(3)
		& 0.18			& 5.5
		& 7
		& 31 			& 2
								\\
$m_{1^1S_0}$, $\triangle m_{2^1S_0 - 1^1S_0}$ 
		& 14-32 		& 1.0(2)
		& 0.53			& 20
		& 13 			& 37
		& 2 
								\\
$m_{1^3S_1}$, $\triangle m_{2^3S_1 - 1^3S_1}$ 
		& 14-32 		& 0.9(2)
		& 0.60			& 22
		& 13 			& 37
		& 2
								\\
$m_{1^1P_1}$, $\triangle m_{2^1P_1 - 1^1P_1}$  
		& 11-32 		& 2.1(2)
		& $10^{-5}$		& $4\times10^{-3}$
		& 9			& 50
		& 2
								\\
$m_{1^3P_0}$, $\triangle m_{2^3P_0 - 1^3P_0}$
		& 10-32 		& 1.5(2)
		& 0.02			& 0.7
		& 16			& 46
		& 2
								\\
$m_{1^3P_1}$, $\triangle m_{2^3P_1 - 1^3P_1}$
		& 10-32 		& 1.3(2)
		& 0.08			& 3.4
		& 17			& 45
		& 2
								\\
\end{tabular}
\end{center}
\caption[Fitting details at $\beta = 6.1$ on a $16^3 \cdot 64$ lattice]{ 
Fitting details at $\beta = 6.1$ on a $16^3 \cdot 64$ lattice.
}
\label{tab:fit_b6.1}
\end{table}



\begin{table}[ptb]
\begin{center}
\begin{tabular}{lcccccccc} 
Meson \& momenta   & $c(0)$
		& fitting			& $\chi^2$ per
		& Q				& goodness
		& dropped			& d.o.f.
		& bin				
								\\ 
     &
		& range				& d.o.f.
		& value				& of fit
		& eigenvalues			& left 
		& size
								\\\hline
$1^1S_0$, $(1,0,0)$ $(2,0,0)$
		& 0.984(3)
		& 17-32 			& 1.3(3)
		& 0.15				& 4.1
		& 14
		& 28	 			& 2
								\\
$1^1S_0$, $(1,1,0)$ $(2,2,0)$
		& 0.979(14)
		& 14-32 			& 1.5(3)
		& 0.06				& 1.6
		& 24			
		& 27	 			& 2
								\\
$1^3S_1$, $(1,0,0)$ $(2,0,0)$
		& 0.986(3)
		& 17-32		 		& 1.3(2)
		& 0.08				& 3.2
		& 1 				& 41
		& 2 
								\\
$1^3P_0$, $(1,0,0)$ $(2,0,0)$
		& 0.973(17)
		& 17-32		 		& 1.2(2)
		& 0.16				& 6.9
		& 0				& 42
		& 2
								\\
\end{tabular}
\end{center}
\caption[Dispersion relation checked 
at $\beta = 6.1$ on a $16^3 \cdot 64$ lattice]{ 
Dispersion relation checked at $\beta = 6.1$ on a $16^3 \cdot 64$ lattice.
As an example on how to read this table, 
the first row gives the effective velocity of light 
$c(\bp = 0)$, fitted from the pseudo-scalar meson $1^1S_0$ with 
momenta 0, $\frac{2\pi}{L}(1,0,0)$ and $\frac{2\pi}{L}(2,0,0)$.
Not listed here due to failed fitting are:
meson $1^3P_1$ and $1^1P_0$ of the $(1,0,0)$ momentum series, 
and all mesons but $1^1S_0$ of the $(1,1,0)$ momentum series.
}
\label{tab:dispersion}
\end{table}


\tabcolsep0.1cm


\begin{table}[ptb]
\begin{center}
\begin{tabular}{cllllcccccc} 
ID  		& \multicolumn{4}{c}{Inputs}
		& \multicolumn{5}{c}{Outputs (masses in MeV)}
		& \# of 
								\\ 
		& $m_0$ & $\nu_t$  & $C_{\rm SW}^s$ & $C_{\rm SW}^t$
		& $c(0)$ 
		& $1{\bar S}$
		& $\triangle S$ 
		& $\triangle P$ 
		& $\triangle SP$ 
		& conf
								\\\hline
0		& 0.51  & 1.01 & 2.178 & 1.3396
		& 0.996(2)  
		& 3070(2) 
		& 79.1(6)
		& 43(3)
		& 455(8)
		& 980
								\\
1		& 0.5355  & 1.01 & 2.178 & 1.3396
		& 1.000(6)  
		& 3125(3) 
		& 78.0(9)
		& 40(5)
		& 467(13)
		& 190
								\\
2		& 0.4845  & 1.01 & 2.178 & 1.3396
		& 1.001(4)  
		& 3017(3) 
		& 82.9(9)
		& 40(6)
		& 463(13)
		& 200
								\\
3		& 0.51  & 1.06 & 2.178 & 1.3396
		& 0.974(5)  
		& 2991(2) 
		& 75.9(9)
		& 30(5)
		& 470(30)
		& 200
								\\
4		& 0.51  & 0.96 & 2.178 & 1.3396
		& 1.035(5)  
		& 3156(3) 
		& 82.4(10)
		& 39(5)
		& 440(15)
		& 230
								\\
5		& 0.51  & 1.01 & 2.396 & 1.3396
		& 1.001(3)  
		& 3033(2) 
		& 92.8(7)
		& 42(4)
		& 463(8)
		& 730
								\\
6		& 0.51  & 1.01 & 2.178 & 1.4736
		& 1.000(2)  
		& 3042(2)
		& 83.3(6)
		& 45(5)
		& 462(8)
		& 470
								\\
\end{tabular}
\end{center}
\caption[Sensitivity to inputs checked 
at $\beta = 5.7$ on an $8^3 \cdot 32$ lattice
]{ 
At $\beta = 5.7, \xi=2$ on an $8^3 \cdot 32$ lattice, 
we check the effects of the inputs on 
the effective velocity of light $c(0)$, 
the 1S spin average meson mass $m(1S)$, 
the $S$-wave mass splitting $\triangle m_{1^3S_1 - 1^1S_0}$,
the $P$-wave mass splitting $\triangle m_{1^3P_1 - 1^3P_0}$, 
and the $S-P$ splitting $\triangle m_{1^1P_1 - 1S}$.
The right-most column gives the number of configurations.
Inputs that are not shown 
are exactly the same as those listed for the $\beta = 5.7$ runs 
in table \ref{tab:basic} and table \ref{tab:sim_cnd}.
}
\label{tab:tuning}
\end{table}



\clearpage
\begin{figure}[ptb]
  \begin{minipage}[t]{120mm}           
    \epsfxsize=120mm                   
    \epsffile{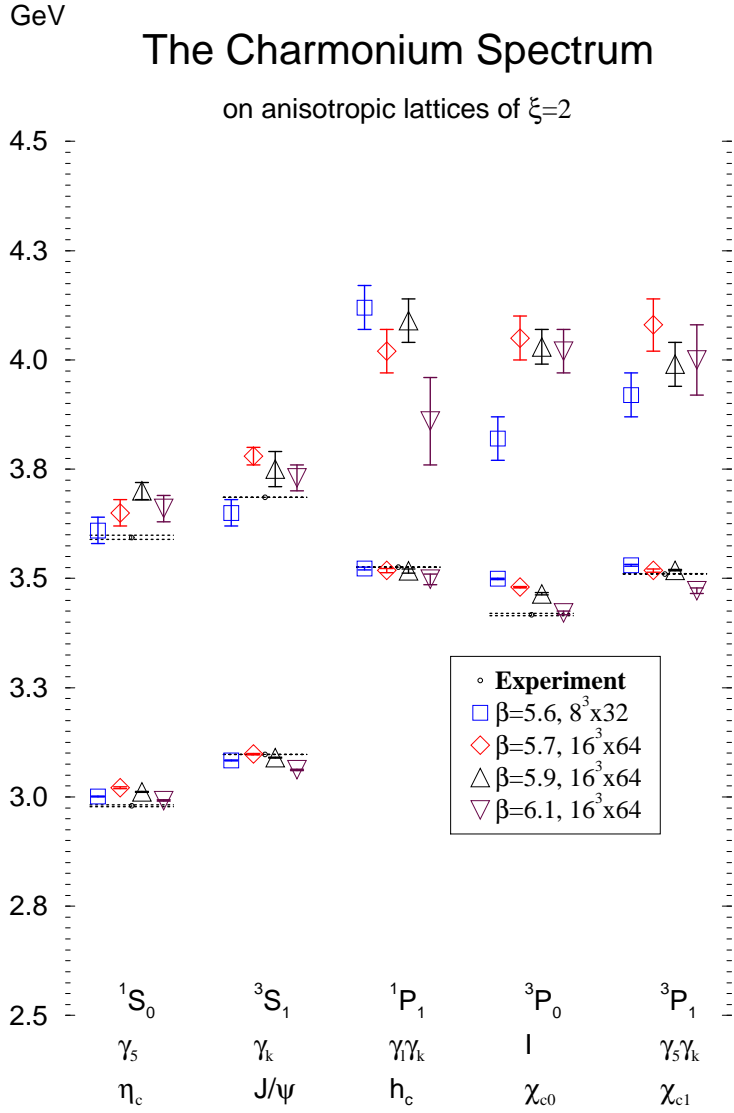}
  \end{minipage}
\caption[Charmonium spectrum compared with experimental values]{
Measured at four values of lattice spacings, the charmonium masses
are plotted on the top of experimental values 
(shown as horizontal lines wherever available).
}
\label{fig:charmonium}
\end{figure}

\clearpage
\begin{figure*}[ptb]

  \begin{minipage}[t]{75mm}
    \epsfxsize=60mm
    \epsffile{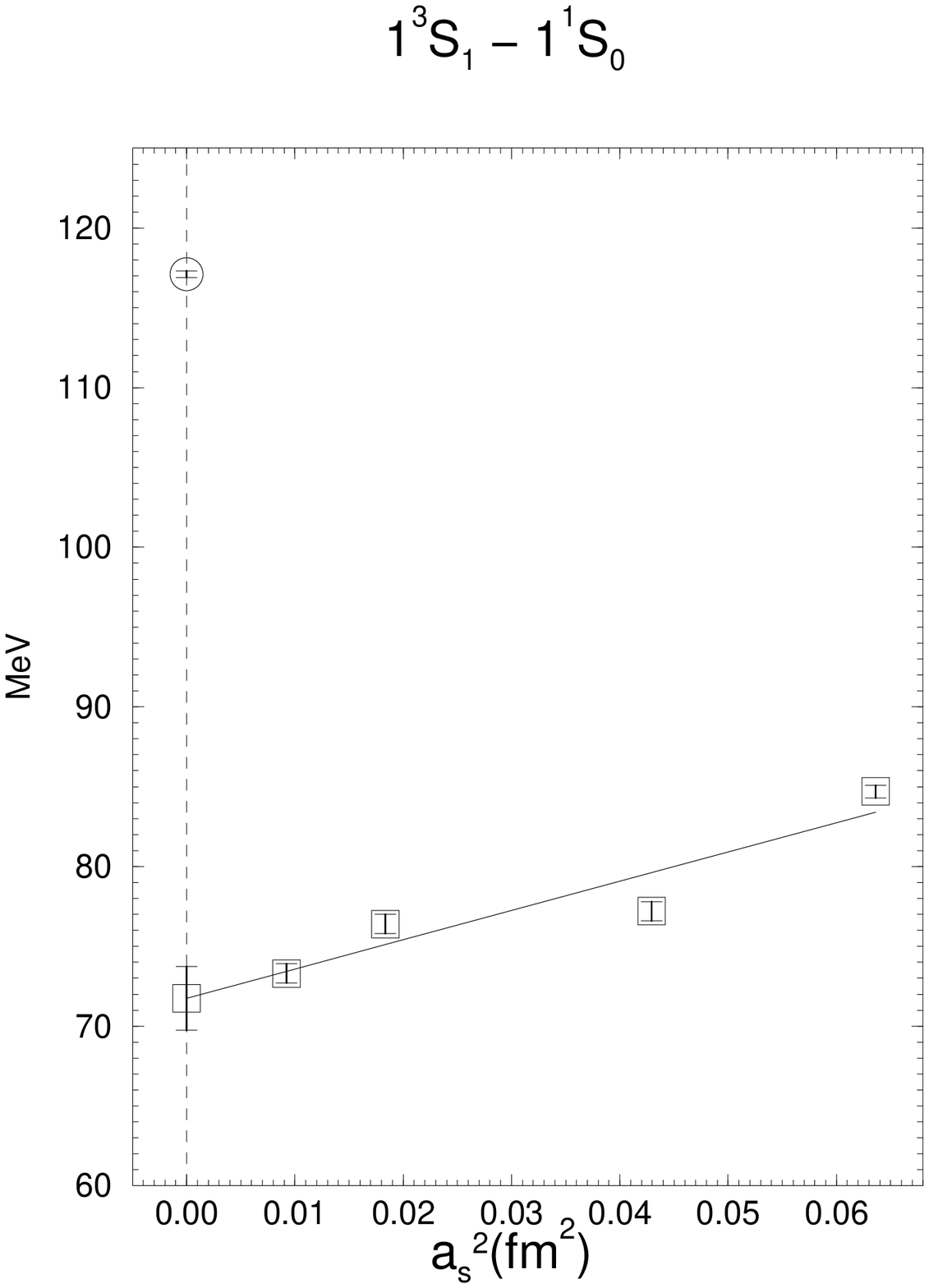}
  \end{minipage}   
  \begin{minipage}[t]{75mm}
    \epsfxsize=60mm
    \epsffile{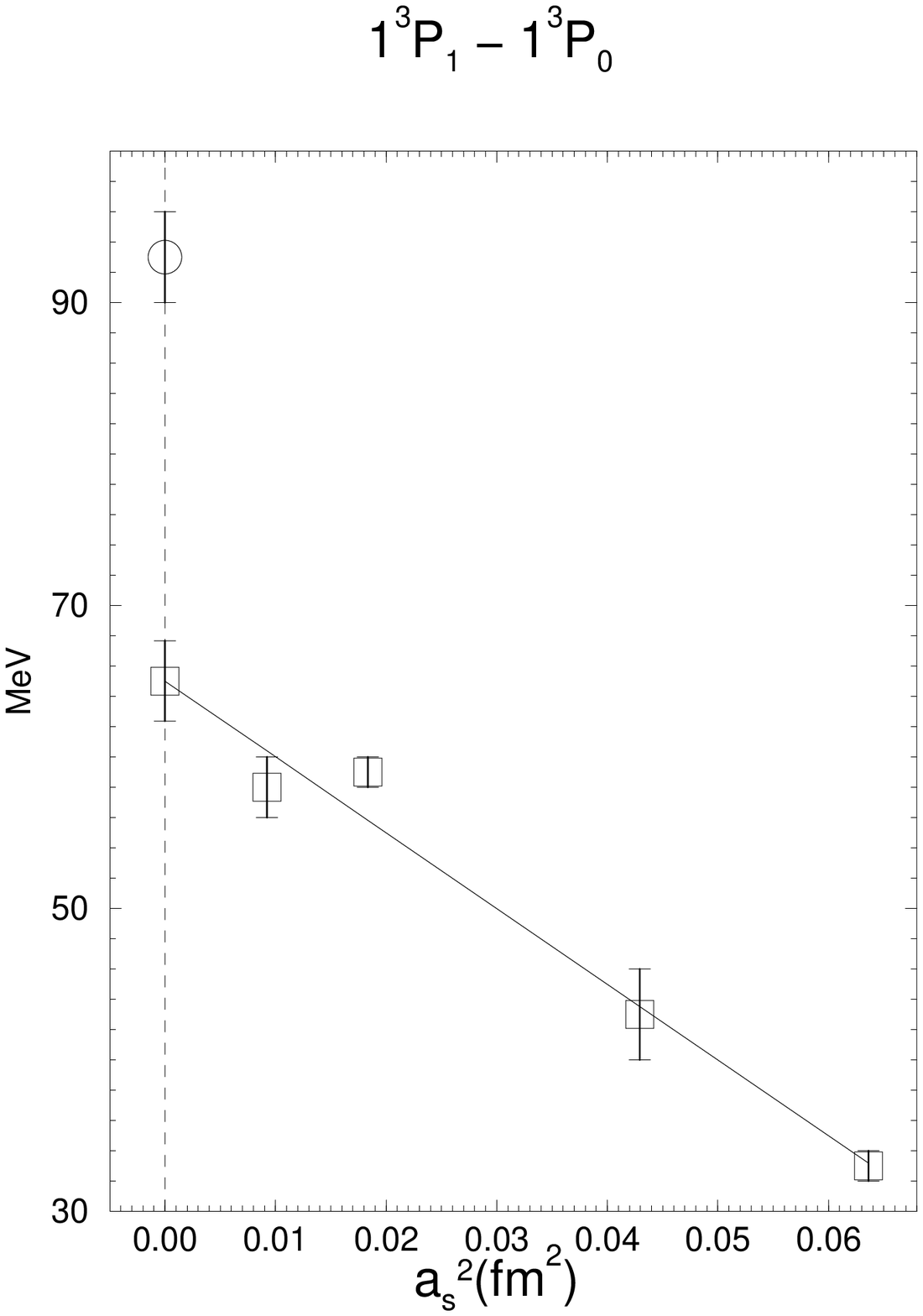}
  \end{minipage}       

  \begin{minipage}[h]{75mm}
    \epsfxsize=60mm
    \epsffile{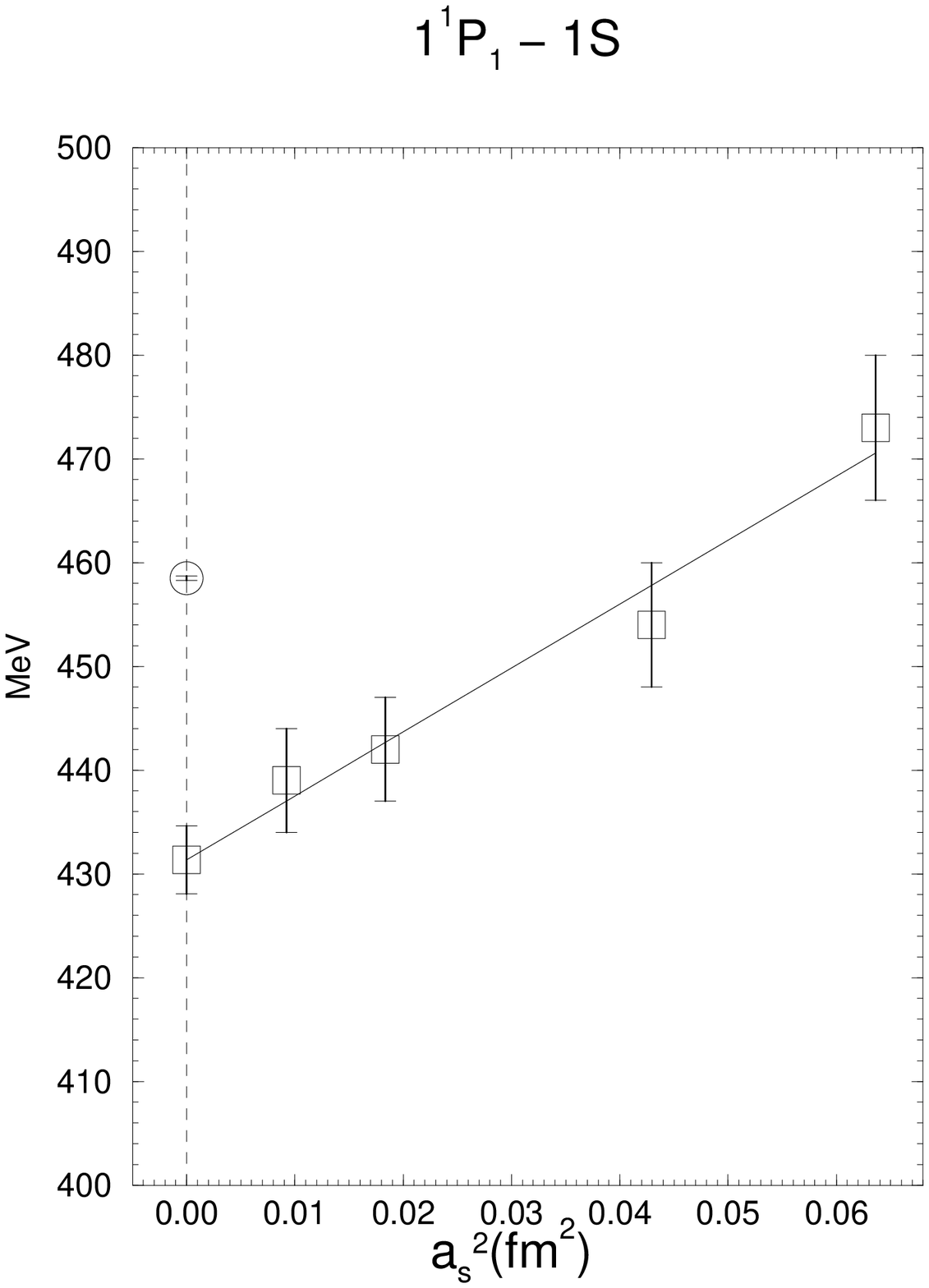}
  \end{minipage}   
  \begin{minipage}[h]{75mm}
    \epsfxsize=60mm
    \epsffile{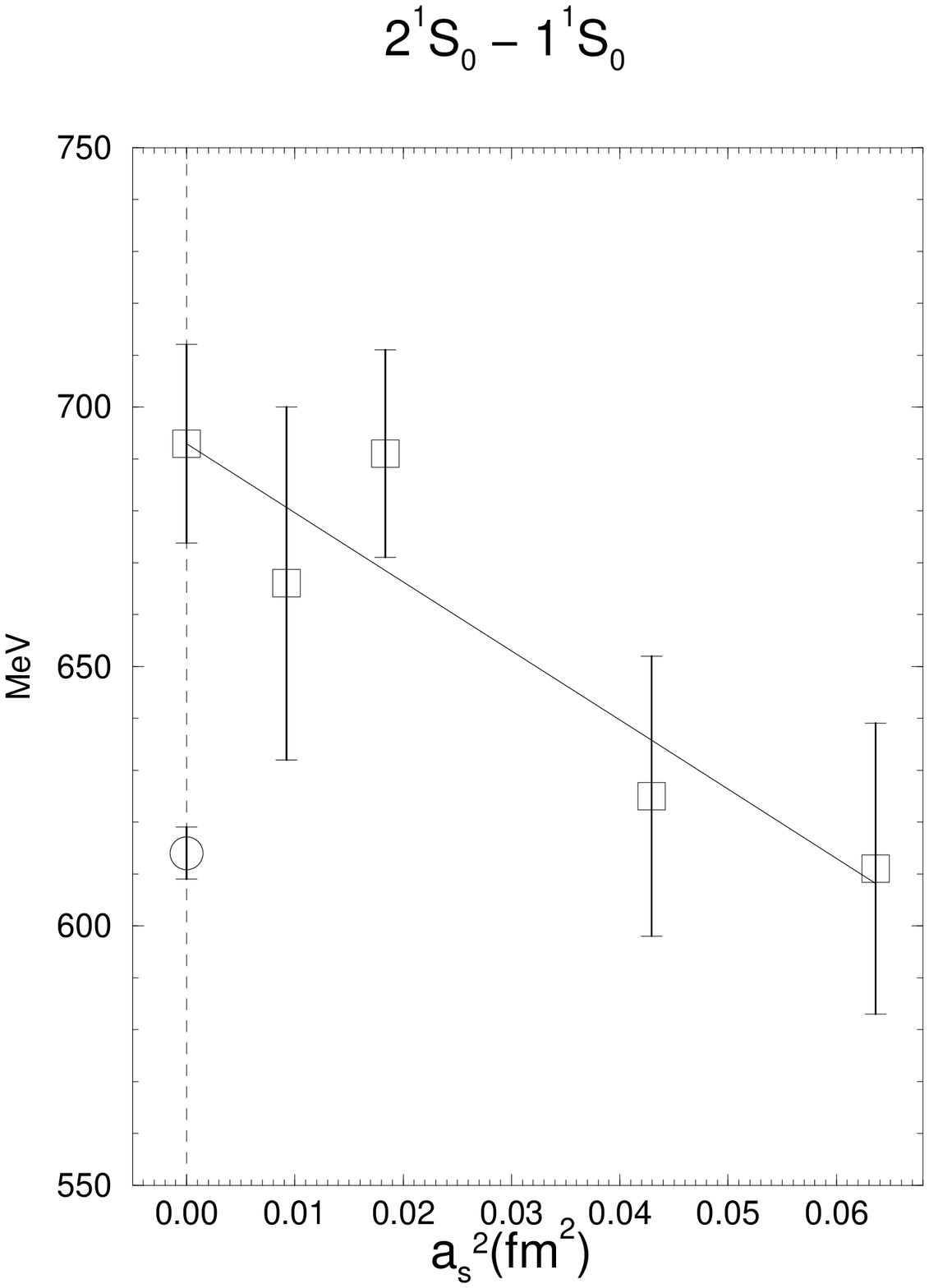}
  \end{minipage}

\caption[Continuum $a_s^2$ extrapolation of mass differences (I)]{
The continuum $a_s^2$ extrapolation of mass differences. 
The four squares correspond to the four $\beta$ values,  
5.6, 5.7, 5.9 and 6.1, at fixed renormalized anisotropy $\xi = 2$. 
The vertical dashed line is there to emphasize the continuum limit. 
Wherever available, a circle on the dashed line is the observed 
mass difference in Nature.
}
\label{fig:splitting0}
\end{figure*}

\clearpage
\begin{figure*}[ptb]

  \begin{minipage}[t]{75mm}
    \epsfxsize=60mm
    \epsffile{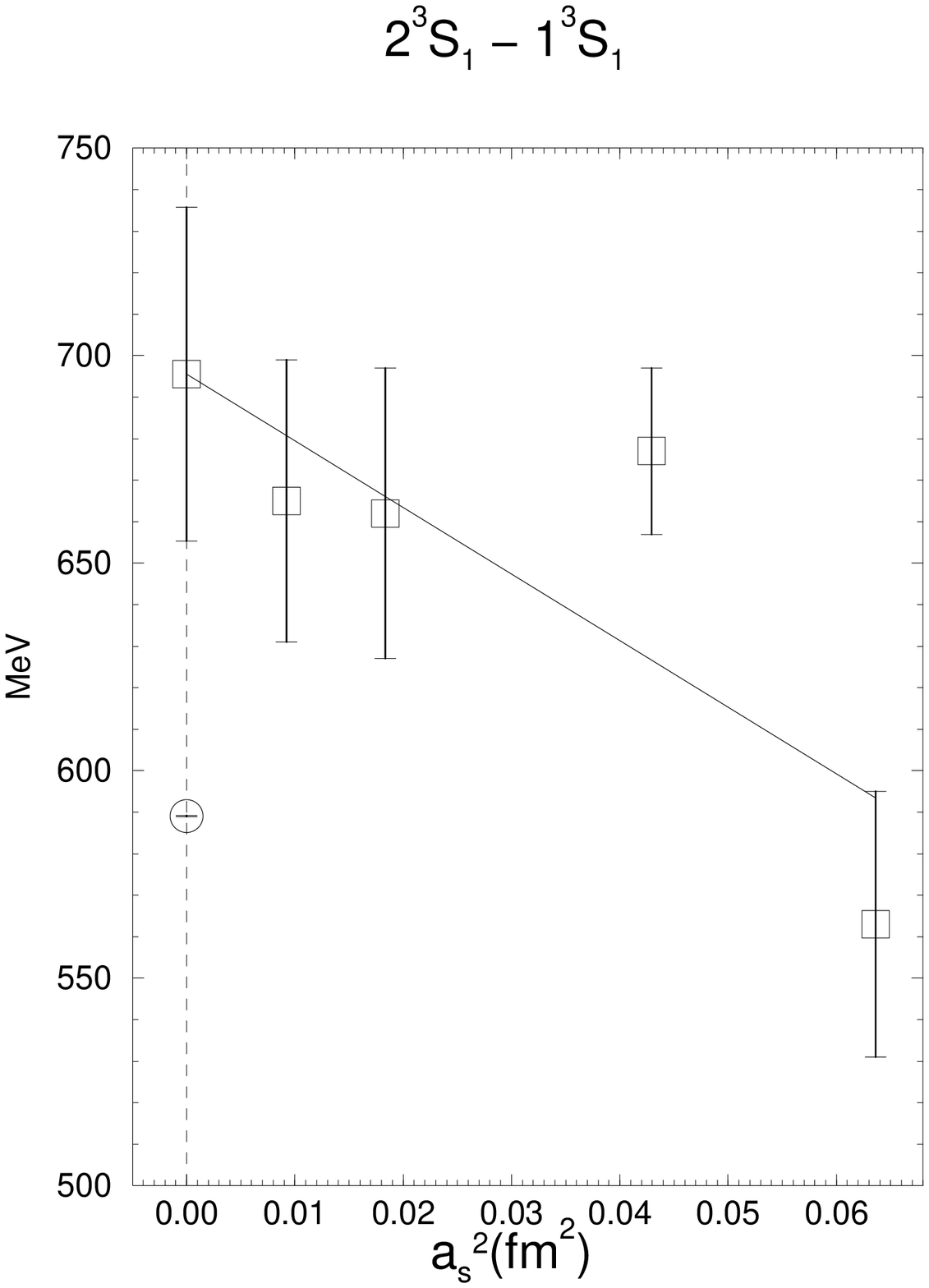}
  \end{minipage}
  \begin{minipage}[t]{75mm}
    \epsfxsize=60mm
    \epsffile{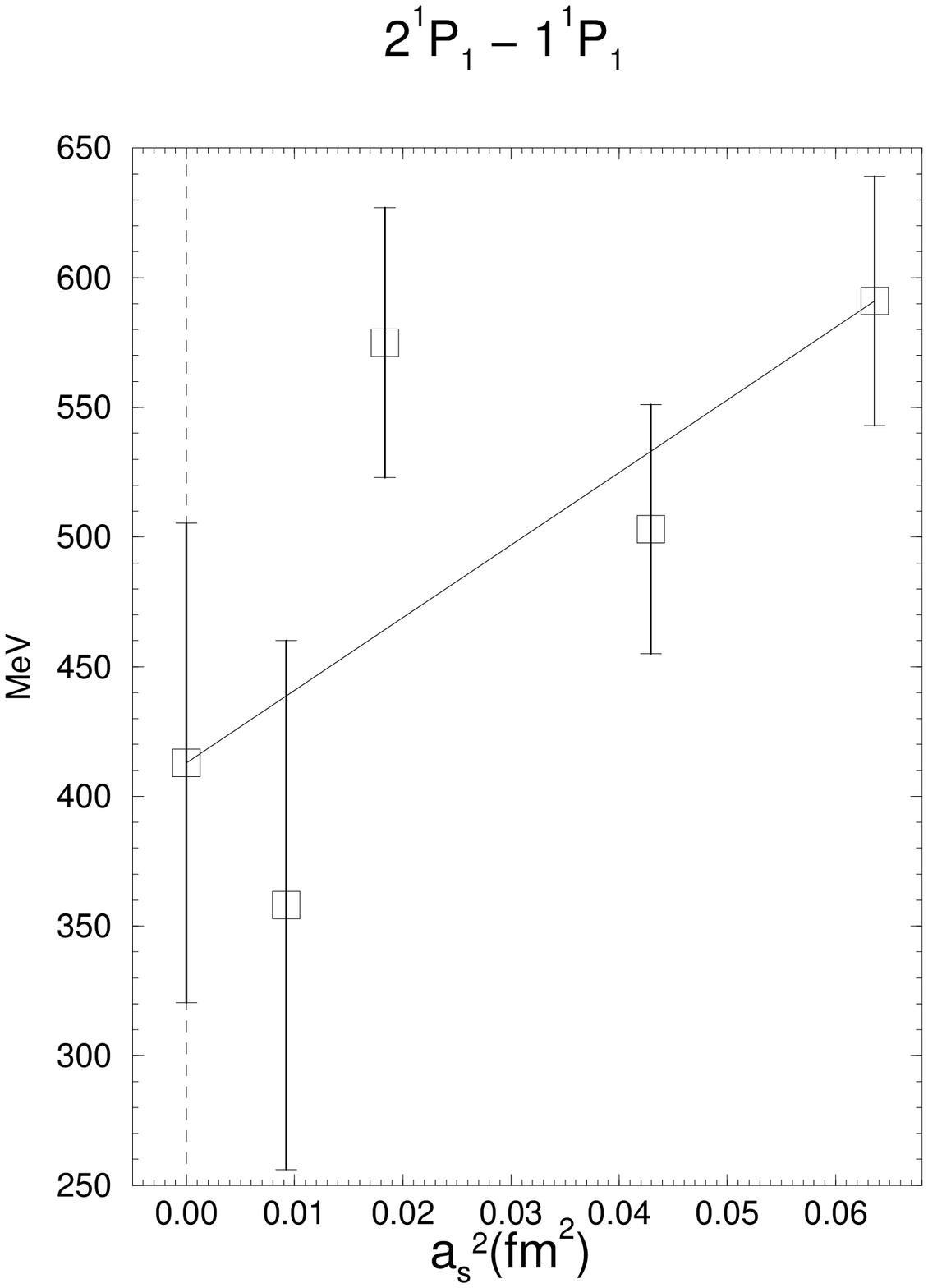}
  \end{minipage}

  \begin{minipage}[h]{75mm}
    \epsfxsize=60mm
    \epsffile{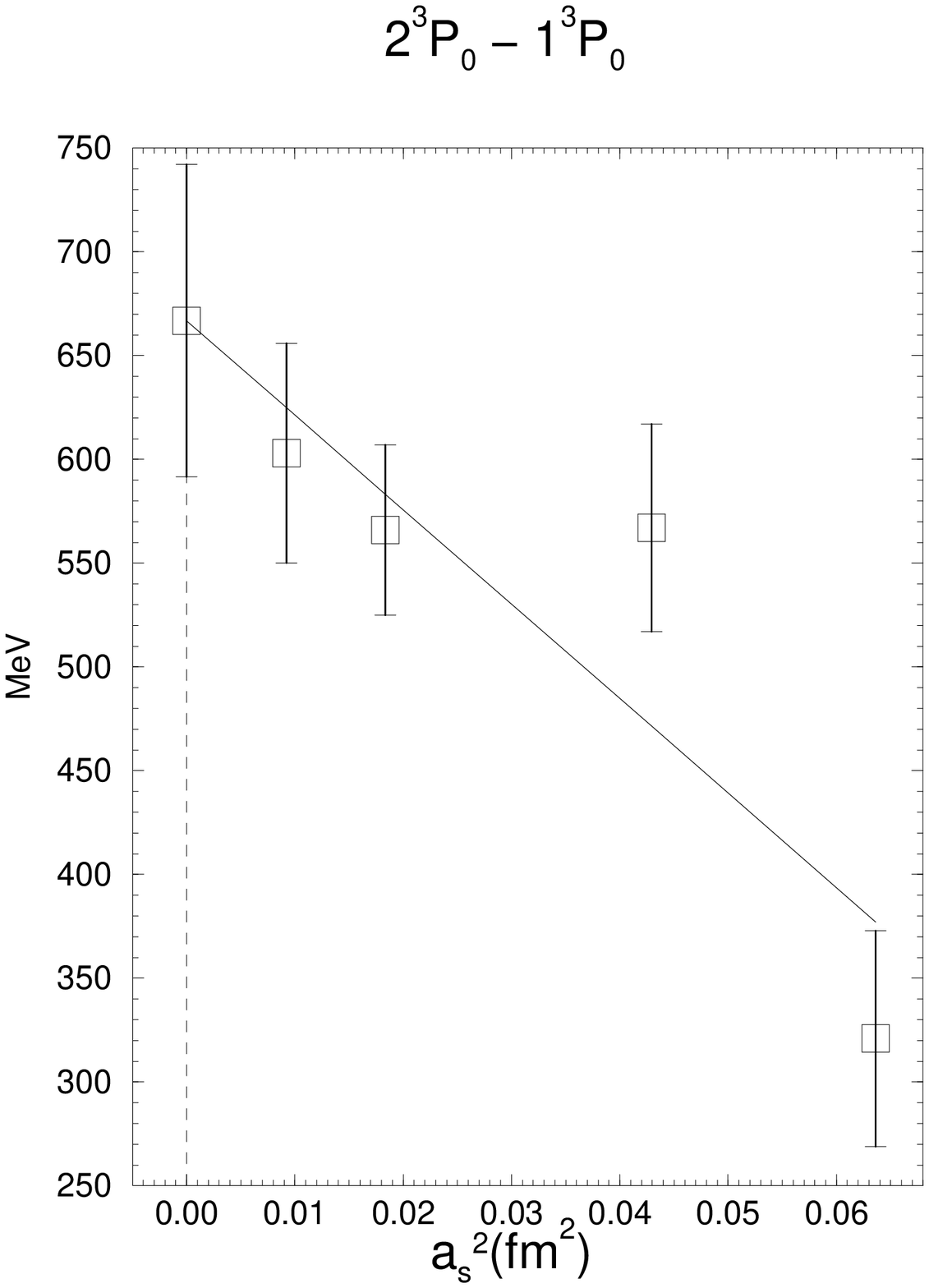}
  \end{minipage}
  \begin{minipage}[h]{75mm}
    \epsfxsize=60mm
    \epsffile{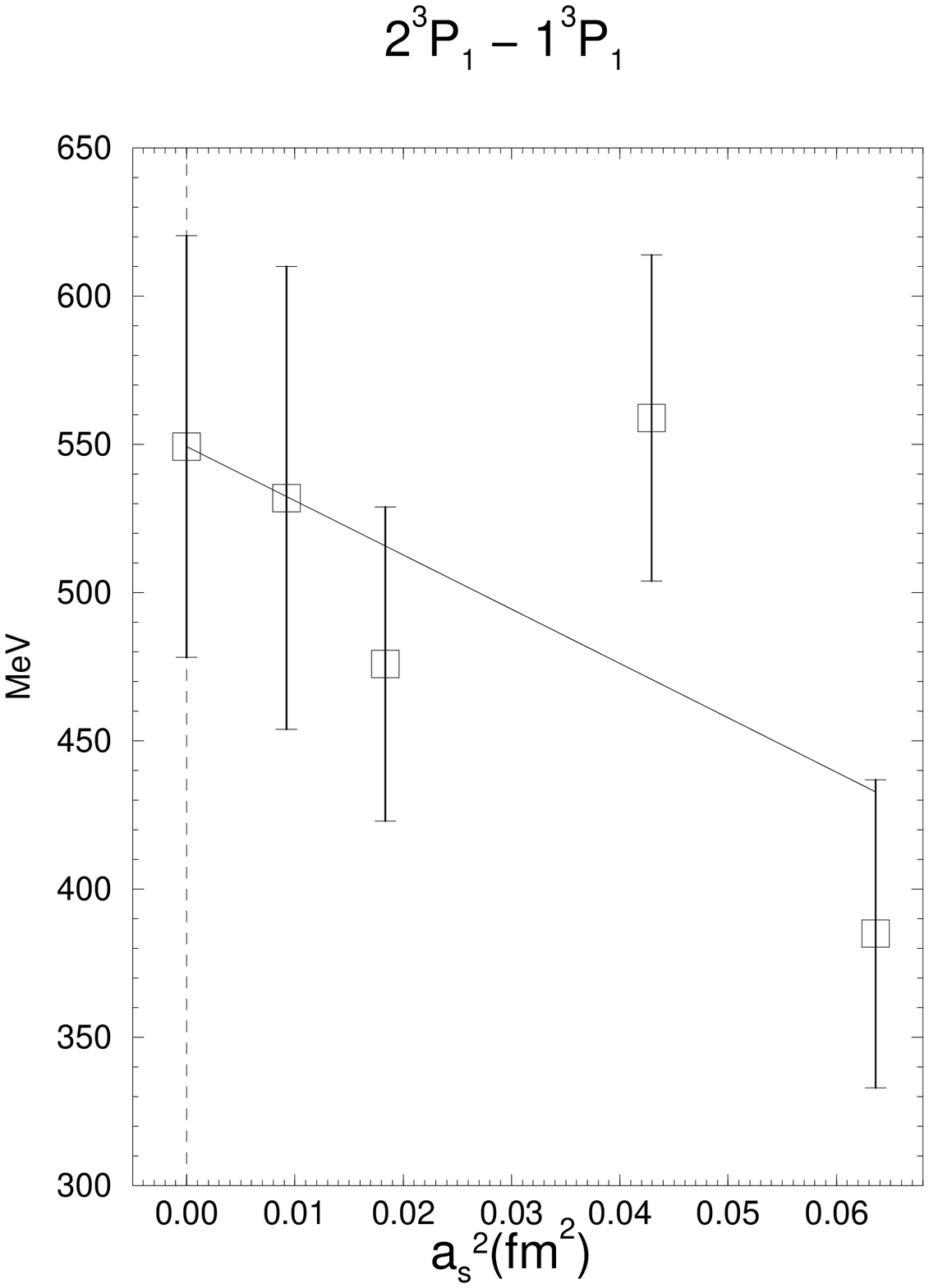}
  \end{minipage}

\caption[Continuum $a_s^2$ extrapolation of mass differences (II)]{
The continuum $a_s^2$ extrapolation of mass differences, 
continued from fig \ref{fig:splitting0}. 
}
\label{fig:splitting1}
\end{figure*}


\clearpage
\begin{figure}[ptb]
  \begin{minipage}[t]{120mm}
    \epsfxsize=120mm
    \epsffile{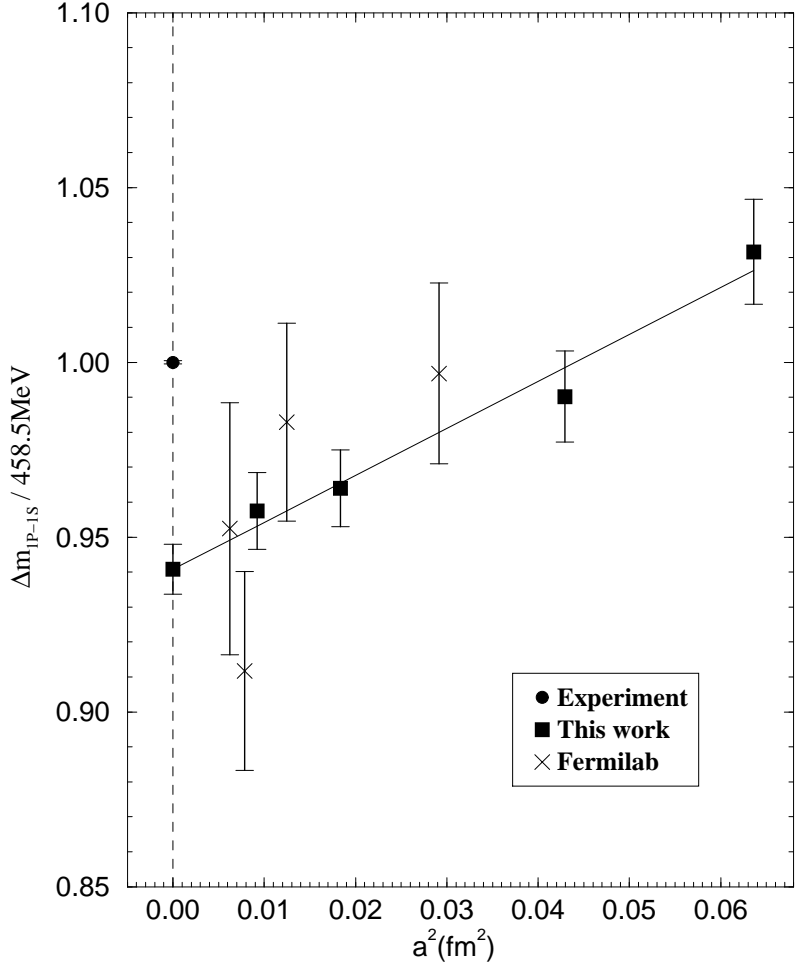}
  \end{minipage}   
\caption[Comparison of the $1P-1S$ splitting]{
The $1^1P_1-1S$ splitting from several approaches is presented
in the form of the ratio 
$\frac{a_{1^1P_1-1S}}{a_{r_0}} =
       \frac{\triangle m_{1^1P_1 - 1S}}{458.5 {\rm  MeV}}$. 
See section \ref{section:ps_splitting} for more details.
The vertical dashed line is to emphasize the continuum limit.
The solid line denotes the $a^2$ continuum extrapolation of our work.
}
\label{fig:SP}
\end{figure}

\clearpage
\begin{figure}[ptb]
  \begin{minipage}[t]{120mm}
    \epsfxsize=120mm
    \epsffile{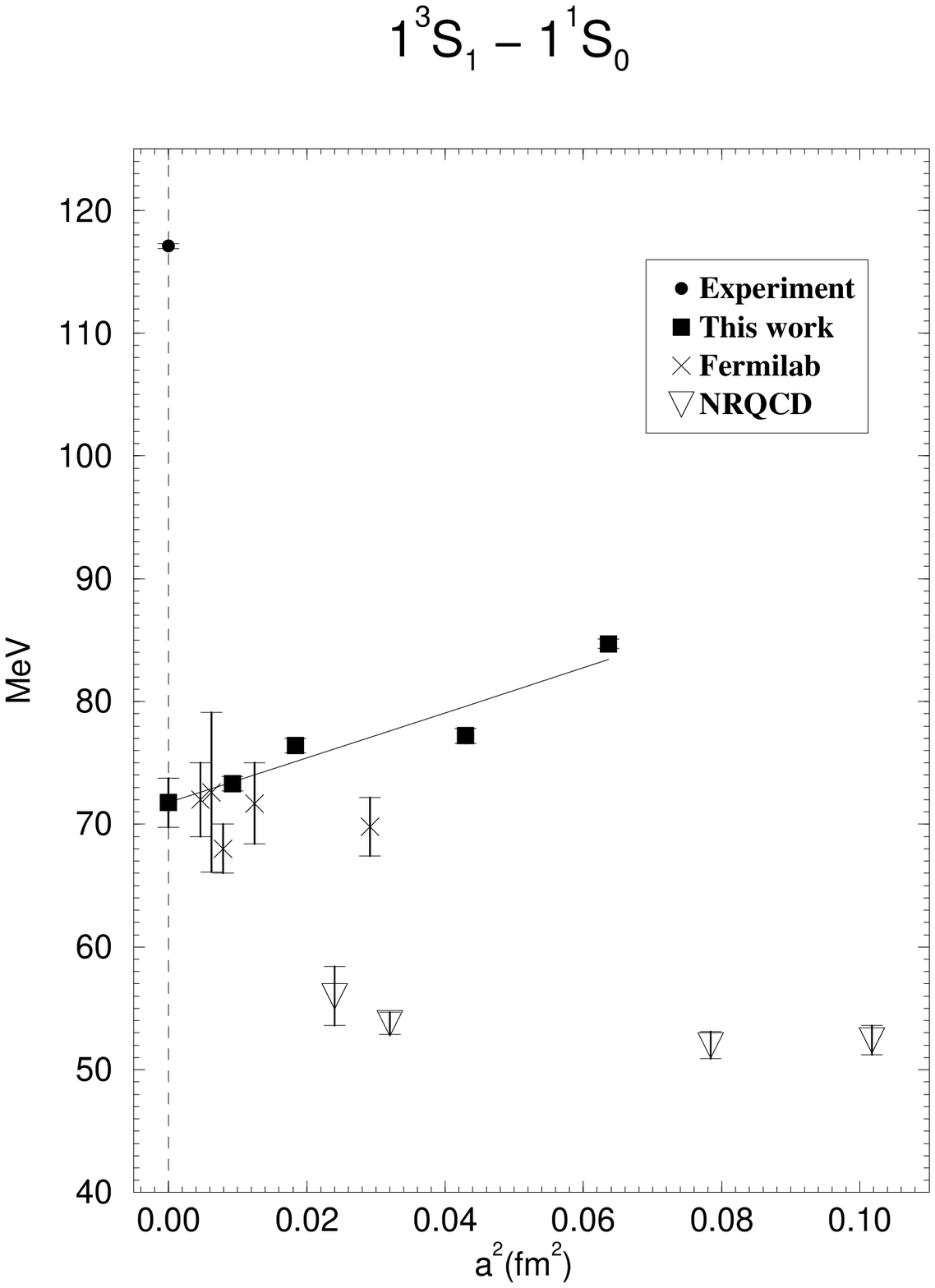}
  \end{minipage}   
\caption[Comparison of the $1^3S_1-1^1S_0$ splitting]{
The $1^3S_1-1^1S_0$ splitting from several approaches.
The vertical dashed line is to emphasize the continuum limit.
The solid line denotes the $a^2$ continuum extrapolation of our work.
}
\label{fig:SS}
\end{figure}

\clearpage
\begin{figure}[ptb]
  \begin{minipage}[t]{120mm}
    \epsfxsize=120mm
    \epsffile{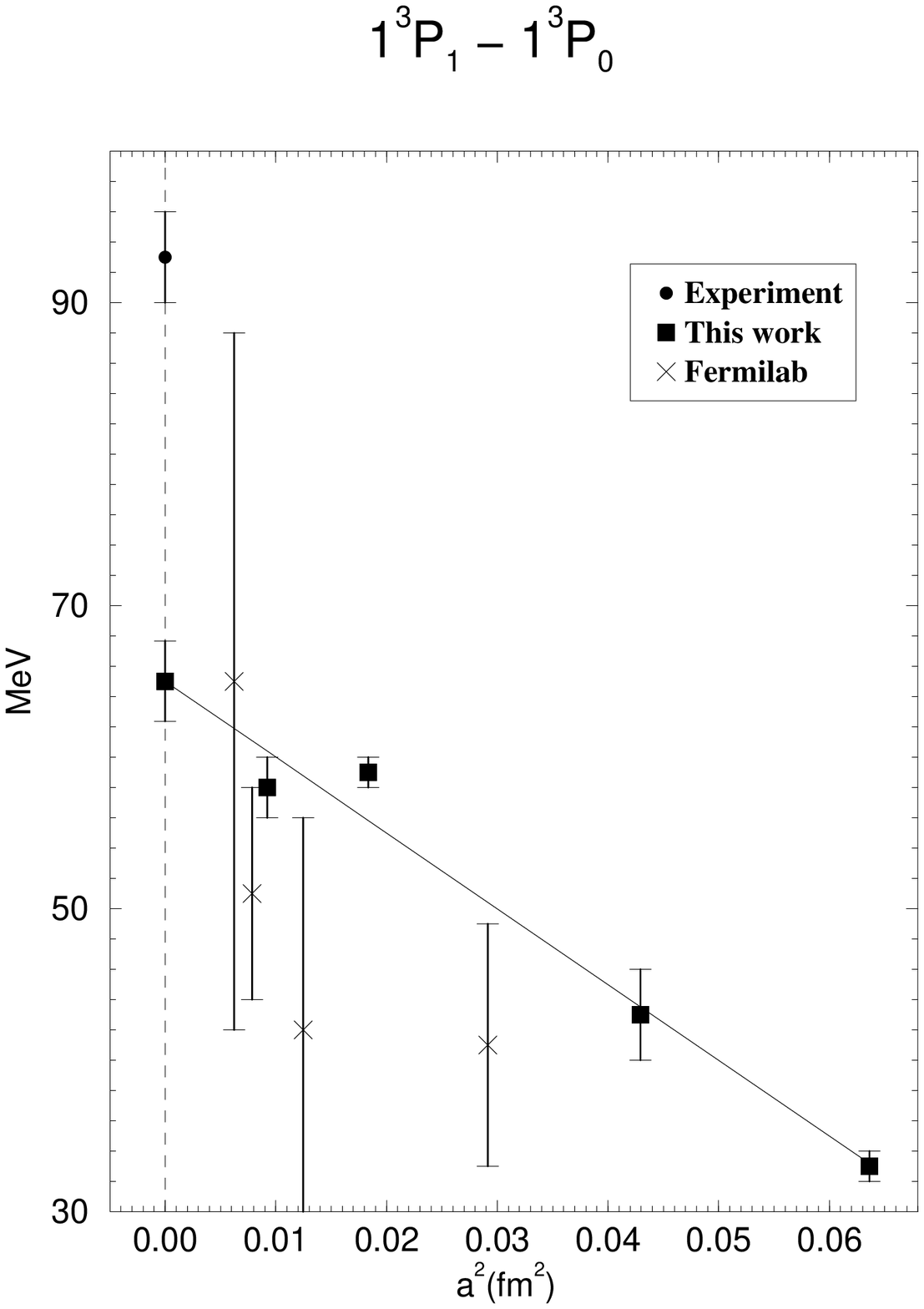}
  \end{minipage}   
\caption[Comparison of the $1^3P_1-1^3P_0$ splitting]{
The $1^3P_1-1^3P_0$ splitting from several approaches.
The vertical dashed line is to emphasize the continuum limit.
The solid line denotes the $a^2$ continuum extrapolation of our work.
}
\label{fig:PP}
\end{figure}

\end{document}